\documentclass[letterpaper,twocolumn,10pt]{article}\usepackage[]{graphicx}\usepackage[]{color}
\makeatletter
\def\maxwidth{ %
  \ifdim\Gin@nat@width>\linewidth
    \linewidth
  \else
    \Gin@nat@width
  \fi
}
\makeatother

\definecolor{fgcolor}{rgb}{0.345, 0.345, 0.345}

\usepackage{framed}
\makeatletter
 {\par\unskip\endMakeFramed%
 \at@end@of@kframe}
\makeatother

\definecolor{shadecolor}{rgb}{.97, .97, .97}
\definecolor{messagecolor}{rgb}{0, 0, 0}
\definecolor{warningcolor}{rgb}{1, 0, 1}
\definecolor{errorcolor}{rgb}{1, 0, 0}

\usepackage{alltt}
\usepackage{usenix}

\usepackage[letterpaper]{geometry}
\usepackage{epsfig,endnotes,url}

\usepackage{xspace}

\usepackage[HideComments]{mycomment}

\usepackage[english]{babel}
\usepackage[latin1]{inputenc}
\usepackage{amsmath,amsthm, amssymb, latexsym}

\usepackage{graphicx}
\usepackage{multirow}
\usepackage{booktabs}

\usepackage{caption,subcaption}

\usepackage{paralist}

\usepackage{balance}

\usepackage{versions}
\excludeversion{DocumentVersionSTAST}
\includeversion{DocumentVersionTR}

\newtheorem{researchquestion}{RQ}

\newcommand{\Zxcvbn}{\textsf{zxcvbn}\xspace}

\newcommand{\const}[1]{\ensuremath{\mathsf{#1}\xspace}}
\newcommand{\vari}[1]{\ensuremath{\mathit{#1}\xspace}}

\newcommand{\CASCAde}{ERC Starting Grant CASCAde (GA n\textsuperscript{o}716980)}

\begin{DocumentVersionSTAST}
\copyrightyear{2017} 
\acmYear{2017} 
\setcopyright{acmcopyright}
\acmConference[STAST2017]{7th International Workshop on Socio-Technical Aspects in Security and Trust}{December 5, 2017}{Orlando, FL, USA}


\end{DocumentVersionSTAST}

\IfFileExists{upquote.sty}{\usepackage{upquote}}{}
\begin{document}




















\newcommand{\fearDescriptivesFear}{
\begin{table}[ht]
\centering
\caption{Descriptives of the Fear Condition} 
\label{tab:fearDescriptivesFear}
\begingroup\footnotesize
\begin{tabular}{rlll}
  \hline
 & \begin{sideways} MeanFearFear \end{sideways} & \begin{sideways} MeanFearJoviality \end{sideways} & \begin{sideways} zxcvbn\_fear\_guesses\_log10 \end{sideways} \\ 
  \hline
mean & 2.91 & 2.35 & 6.39 \\ 
  std.dev & 0.68 & 0.88 & 2.89 \\ 
   \hline
\end{tabular}
\endgroup
\end{table}
}

\newcommand{\fearDescriptivesHappiness}{
\begin{table}[ht]
\centering
\caption{Descriptives of the Happiness Condition} 
\label{tab:fearDescriptivesHappiness}
\begingroup\footnotesize
\begin{tabular}{rlll}
  \hline
 & \begin{sideways} MeanHappyFear \end{sideways} & \begin{sideways} MeanHappyJoviality \end{sideways} & \begin{sideways} zxcvbn\_happy\_guesses\_log10 \end{sideways} \\ 
  \hline
mean & 1.12 & 3.47 &  6.74 \\ 
  std.dev & 0.17 & 0.83 &  3.28 \\ 
   \hline
\end{tabular}
\endgroup
\end{table}
}


\newcommand{\boxplotFearMC}{
\begin{figure}[htb]

\includegraphics[width=\maxwidth]{figure/boxplot_fear-1} 
\caption{Boxplot by condition, color showing the measure.}
\label{fig:boxplotFearMC}
\end{figure}
}

\newcommand{\densityFearMC}{
\begin{figure}[htb]

\includegraphics[width=\maxwidth]{figure/density_fear-1} 
\caption{Density of PANAS-X \textsf{fear} across conditions.}
\label{fig:densityFearMC}
\end{figure}
}

\newcommand{\densityJovMC}{
\begin{figure}[htb]

\includegraphics[width=\maxwidth]{figure/density_jov-1} 
\caption{Density of PANAS-X joviality across conditions.}
\label{fig:densityJovMC}
\end{figure}
}




\newcommand{\esFear}{
\begin{figure}[htb]
\vspace{-2.5cm}
\includegraphics[width=\maxwidth]{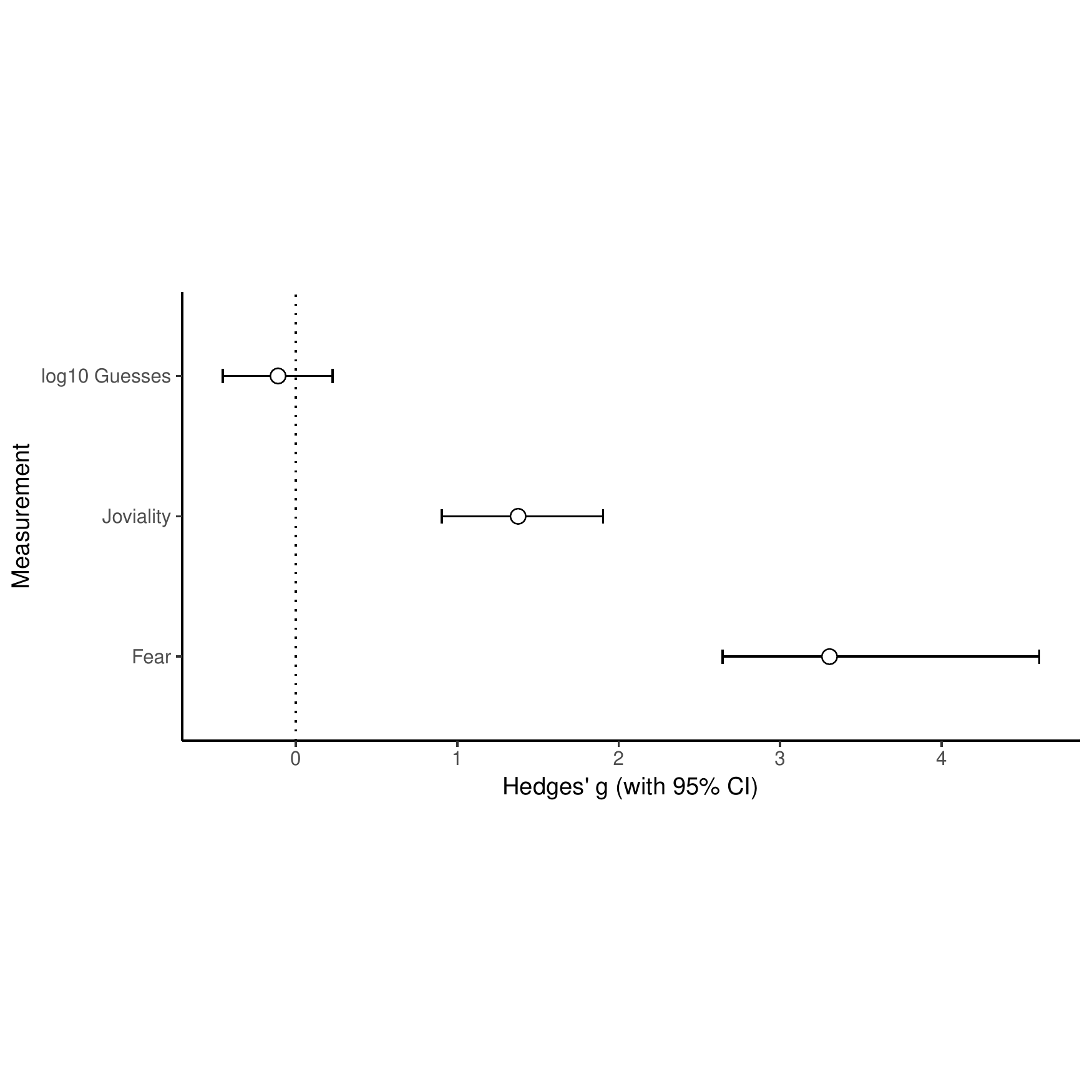} 
\vspace{-2.5cm}\caption{Comparison of the effects of the manipulations vis-`a-vis of the effect of fear on password choice. (Hedges' $g_\const{av}$, 95\% Confidence Intervals).}
\label{fig:esFear}
\end{figure}
}

%
%

\newcommand{\corrgramFear}{
\begin{figure}[htb]

\includegraphics[width=\maxwidth]{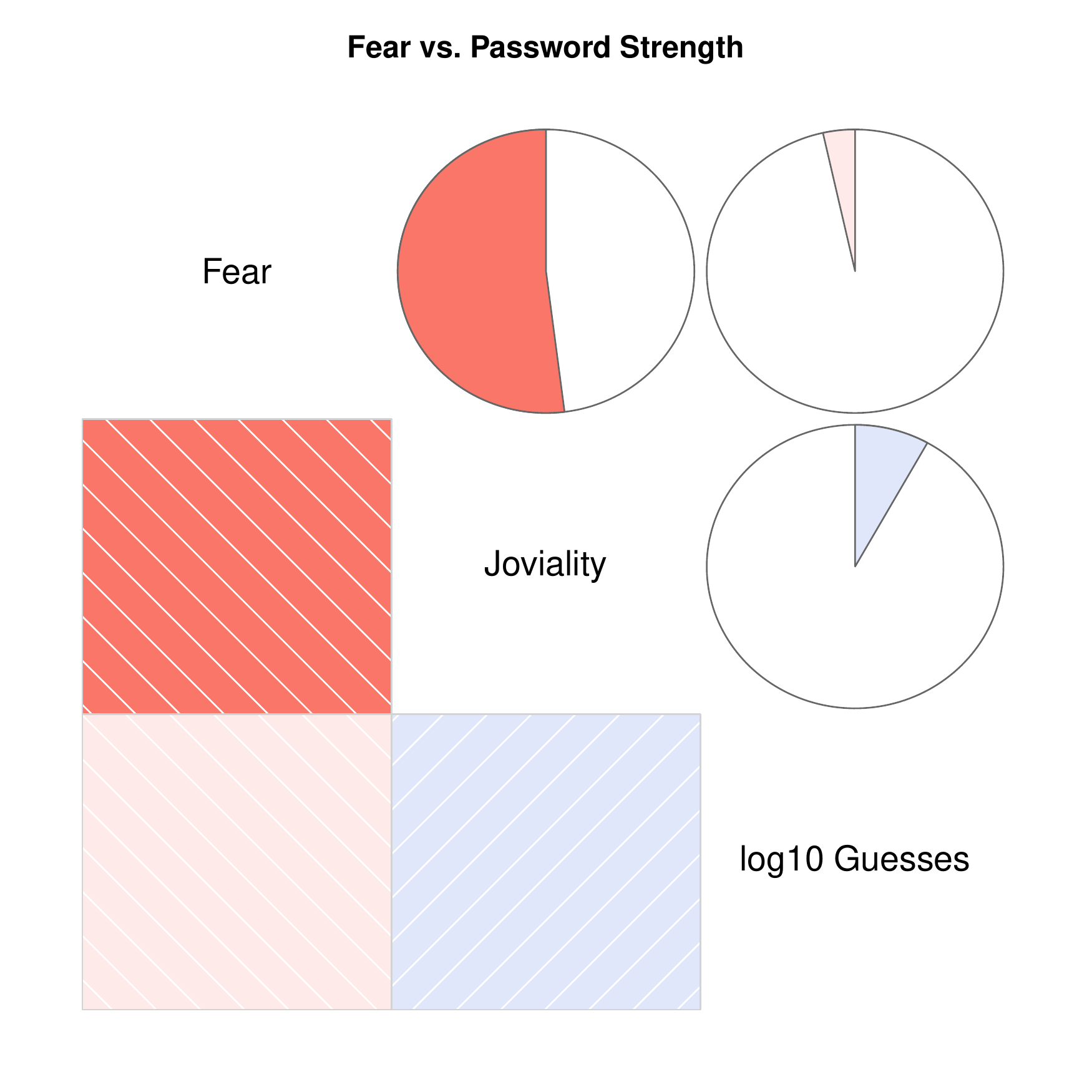} 
\caption{Corrgram of the variables irrespective of conditions.}
\label{fig:corrgramFear}
\end{figure}
}









\newcommand{\stressDescriptivesStress}{
\begin{table}[ht]
\centering
\caption{Descriptives of the Stress Condition} 
\label{tab:stressDescriptivesStress}
\begingroup\footnotesize
\begin{tabular}{rllll}
  \hline
 & \begin{sideways} exp\_total \end{sideways} & \begin{sideways} exp\_distress \end{sideways} & \begin{sideways} exp\_stai\_y\_1 \end{sideways} & \begin{sideways} exp\_zxcvbn\_log10 \end{sideways} \\ 
  \hline
mean & 71.50 & 16.78 &  38.50 & 7.96 \\ 
  std.dev &  8.43 &  5.10 &  10.61 & 2.49 \\ 
   \hline
\end{tabular}
\endgroup
\end{table}
}

\newcommand{\stressDescriptivesCtrl}{
\begin{table}[ht]
\centering
\caption{Descriptives of the Control Condition} 
\label{tab:stressDescriptivesCtrl}
\begingroup\footnotesize
\begin{tabular}{rllll}
  \hline
 & \begin{sideways} ctrl\_total \end{sideways} & \begin{sideways} ctrl\_distress \end{sideways} & \begin{sideways} ctrl\_stai\_y\_1 \end{sideways} & \begin{sideways} ctrl\_zxcvbn\_log10 \end{sideways} \\ 
  \hline
mean & 65.58 & 12.76 & 31.58 & 7.95 \\ 
  std.dev &  9.15 &  4.04 &  7.07 & 2.35 \\ 
   \hline
\end{tabular}
\endgroup
\end{table}
}


\newcommand{\boxplotStressMC}{
\begin{figure}[htb]
\vspace{-2.5cm}
\includegraphics[width=\maxwidth]{figure/boxplot_stress-1} 
\vspace{-2.5cm}\caption{Boxplot by condition, color showing the measure.}
\label{fig:boxplotStressMC}
\end{figure}
}

\newcommand{\densityStressStaiMC}{
\begin{figure}[htb]

\includegraphics[width=\maxwidth]{figure/density_stress_stai-1} 
\caption{Density of STAI State Anxiety across conditions.}
\label{fig:densityStressStaiMC}
\end{figure}
}

\newcommand{\densityStressSssqMC}{
\begin{figure}[htb]

\includegraphics[width=\maxwidth]{figure/density_stress_sssq-1} 
\caption{Density of SSSQ Total Stress across conditions.}
\label{fig:densityStressSssqMC}
\end{figure}
}

\newcommand{\densityStressDistressMC}{
\begin{figure}[htb]

\includegraphics[width=\maxwidth]{figure/density_stress_distress-1} 
\caption{Density of SSSQ Distress across conditions.}
\label{fig:densityStressDistressMC}
\end{figure}
}




\newcommand{\interactConditionTLX}{
\begin{figure}[htb]

\includegraphics[width=\maxwidth]{figure/tlx_interaction_plots-1} 
\caption{Interaction plot on Condition and TLX level.}
\label{fig:interactConditionTLX}
\end{figure}
}
\newcommand{\interactTLXCondition}{
\begin{figure}[htb]

\includegraphics[width=\maxwidth]{figure/tlx_interaction_plots-2} 
\caption{Interaction plot on TLX level and Condition.}
\label{fig:interactTLXCondition}
\end{figure}
}
\newcommand{\interactConditionTLXMental}{
\begin{figure}[htb]
\centering
\begin{subfigure}{0.8\columnwidth}
\centering

\includegraphics[width=\maxwidth]{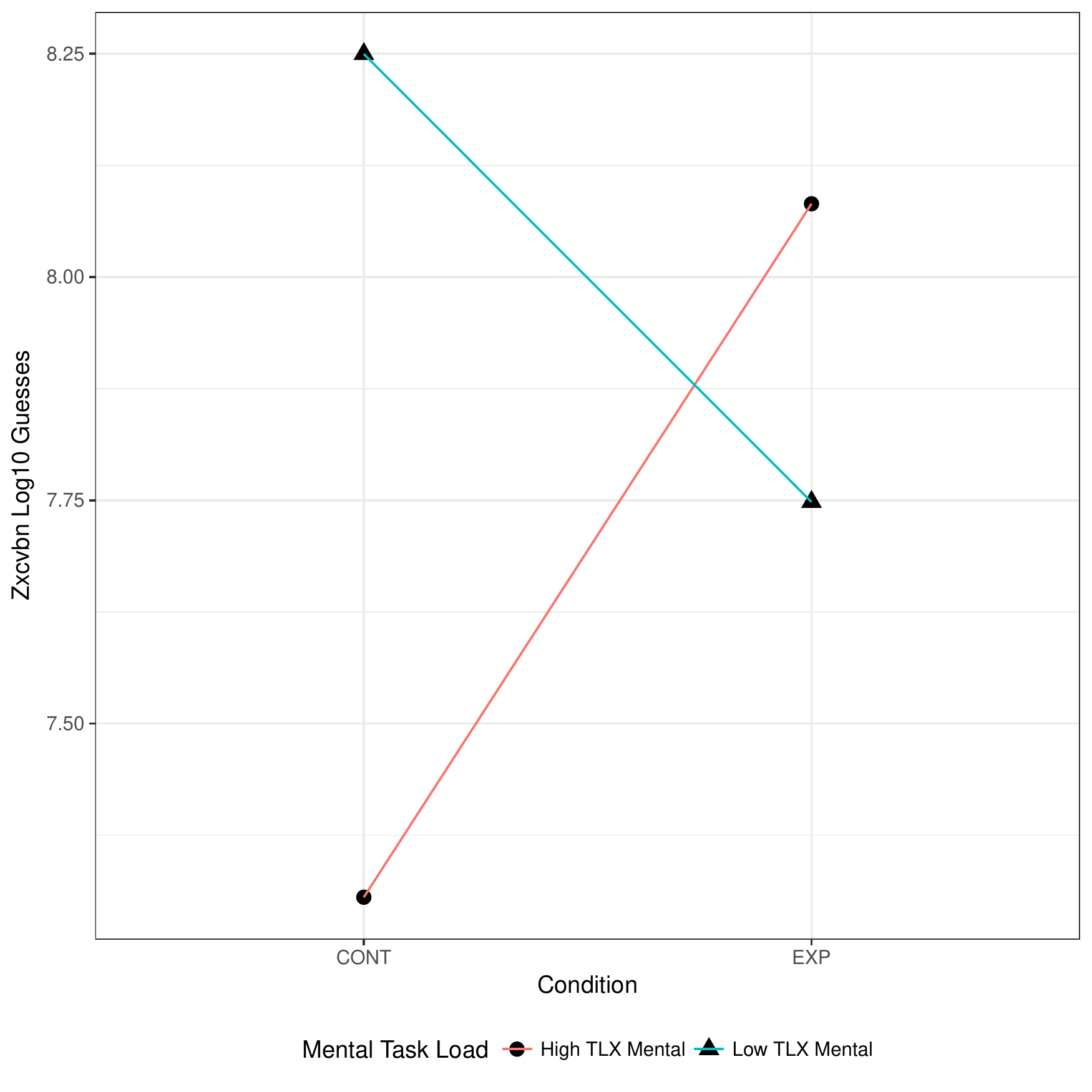} 
\end{subfigure}
\caption{Interaction TLX Mental level by Condition.}
\label{fig:interactConditionTLXMental}
\end{figure}
}
\newcommand{\interactSubFigConditionTLX}{
\begin{figure*}[htb]
\begin{subfigure}{0.49\textwidth}

\includegraphics[width=\maxwidth]{figure/tlx_interaction_plots-4} 
\subcaption{TLX level by Condition.}
\label{fig:interactConditionTLX}
\end{subfigure}~
\begin{subfigure}{0.49\textwidth}

\includegraphics[width=\maxwidth]{figure/tlx_interaction_plots-5} 
\subcaption{Condition by TLX Level.}
\label{fig:interactTLXCondition}
\end{subfigure}
\caption{Interaction plots on Condition and TLX level.}
\label{fig:interactSubFigConditionTLX}
\end{figure*}
}
\newcommand{\interactSubFigConditionTLXMental}{
\begin{figure*}[htb]
\begin{subfigure}{0.49\textwidth}

\includegraphics[width=\maxwidth]{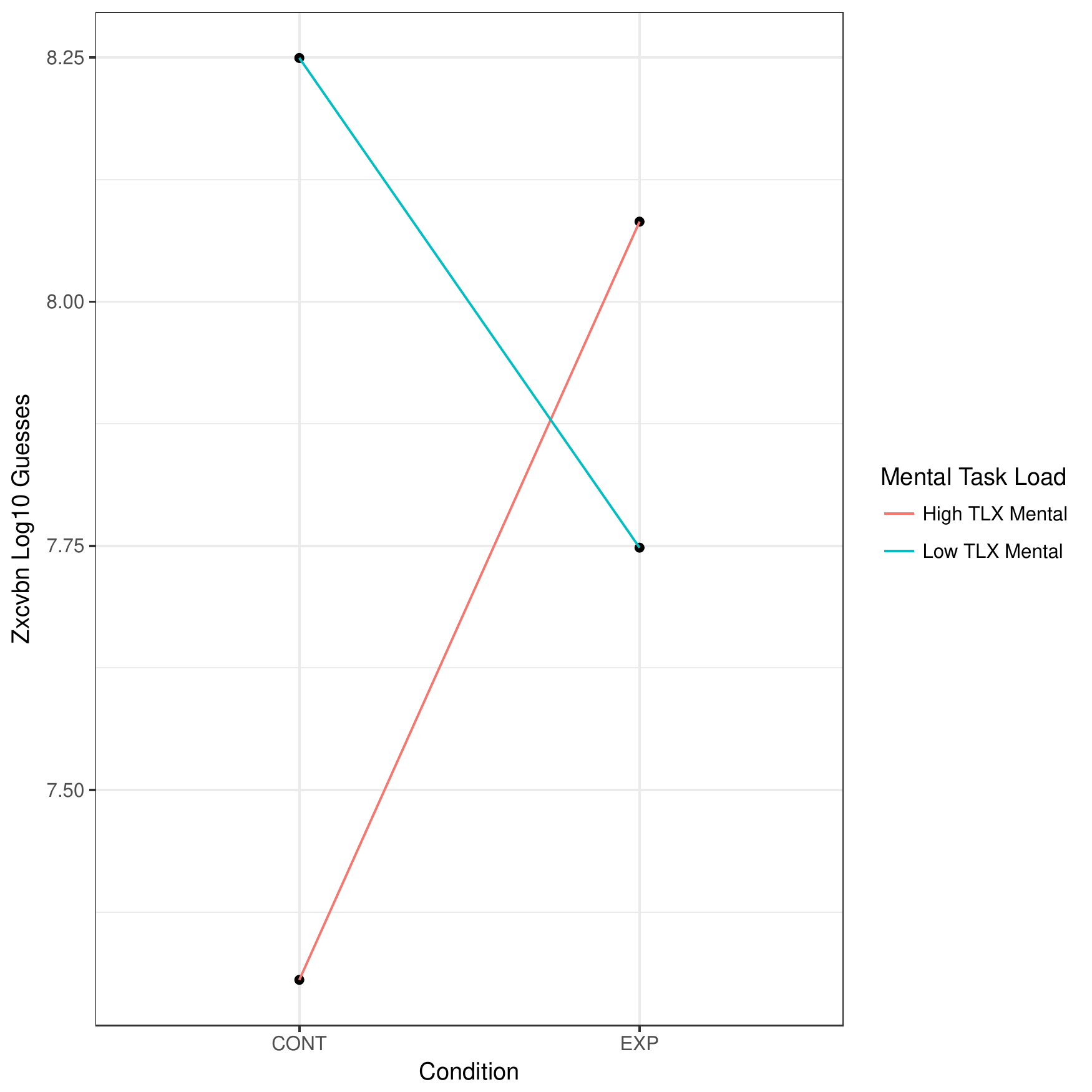} 
\subcaption{TLX Mental level by Condition.}
\label{fig:interactConditionTLXMental}
\end{subfigure}~
\begin{subfigure}{0.49\textwidth}

\includegraphics[width=\maxwidth]{figure/tlx_interaction_plots-7} 
\subcaption{Condition by TLX Mental Level.}
\label{fig:interactTLXMentalCondition}
\end{subfigure}
\caption{Interaction plots on Condition and TLX Mental level.}
\label{fig:interactSubFigConditionTLXMental}
\end{figure*}
}

\newcommand{\interactModelCoef}{
\begin{figure}[htb]
\vspace{-2.5cm}
\includegraphics[width=\maxwidth]{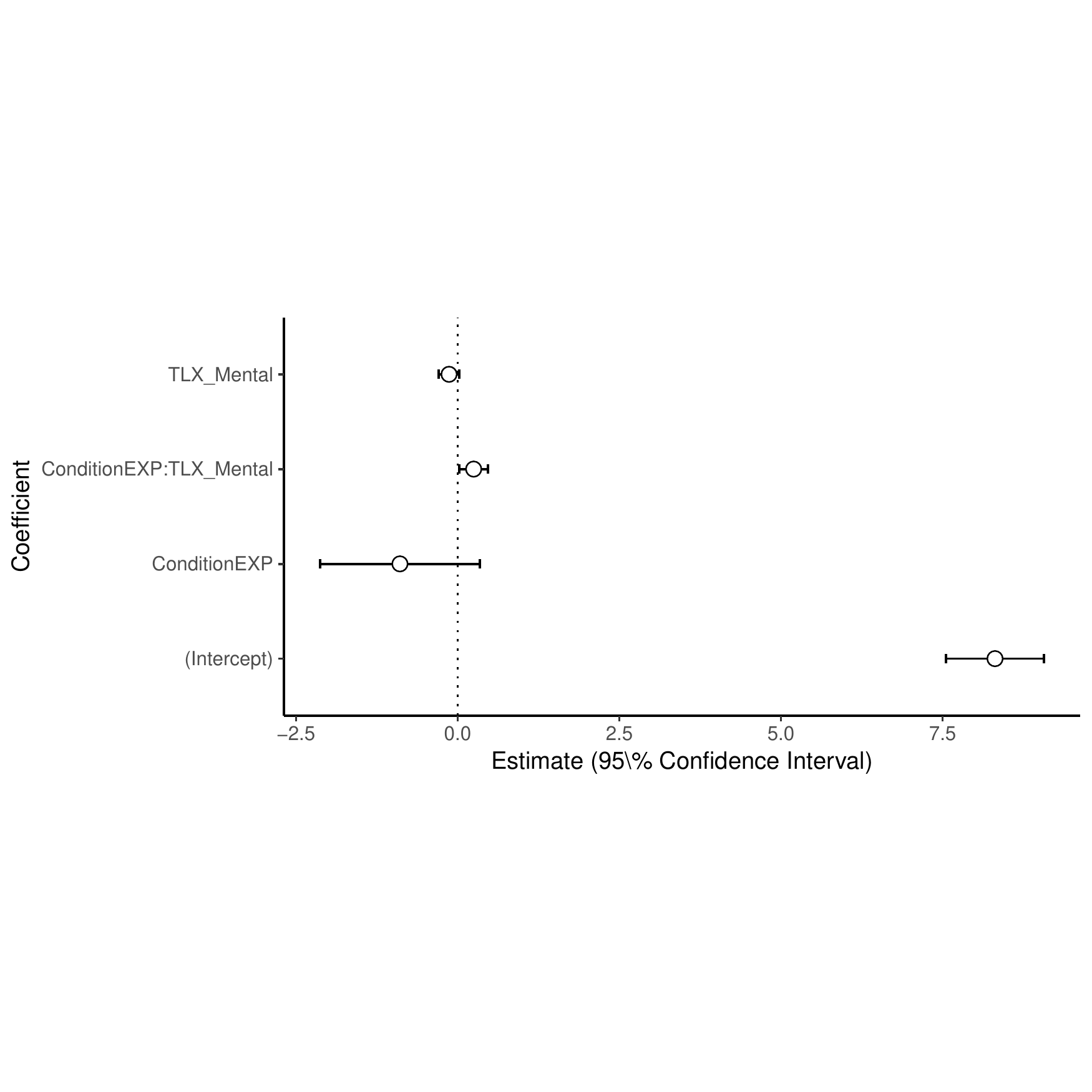} 
\vspace{-2.5cm}\caption{Interaction Model Coefficient Estimates.}
\label{fig:interactModelCoef}
\end{figure}
}

\newcommand{\interactModelCoefReuse}{
\begin{figure}[htb]
\vspace{-2.5cm}
\includegraphics[width=\maxwidth]{figure/tlx_intercept_model_reuse_coefficients-1} 
\vspace{-2.5cm}\caption{Interaction Model Coefficient Estimates.}
\label{fig:interactModelCoefReuse}
\end{figure}
}

\newcommand{\assumLME}{
\begin{figure*}[htb]

\includegraphics[width=\maxwidth]{figure/assumptions_lme_model_stress_interaction-1} 
\caption{Overview of LME assumption plots.}
\label{fig:assumLME}
\end{figure*}
}
\newcommand{\assumLMEhist}{
\begin{figure}[htb]

\includegraphics[width=\maxwidth]{figure/assumptions_lme_model_stress_interaction-2} 
\caption{Histogram of Pearson residuals.}
\label{fig:assumLMEhist}
\end{figure}
}
\newcommand{\assumLMEqqplot}{
\begin{figure}[htb]

\includegraphics[width=\maxwidth]{figure/assumptions_lme_model_stress_interaction-3} 
\caption{QQPlot of Pearson residuals.}
\label{fig:assumLMEqqplot}
\end{figure}
}
\newcommand{\assumLMEscatter}{
\begin{figure}[htb]

\includegraphics[width=\maxwidth]{figure/assumptions_lme_model_stress_interaction-4} 
\caption{Scatterplot of fitted values vs. residuals.}
\label{fig:assumLMEscatter}
\end{figure}
}
\newcommand{\assumFitLME}{
\begin{figure}[htb]

\includegraphics[width=\maxwidth]{figure/assumptions_lme_model_stress_interaction-5} 
\caption{Comparison of fitted and observed values.}
\label{fig:assumFitLME}
\end{figure}
}
\newcommand{\assumFitLMEplain}{
\begin{figure}[htb]
\centering
\begin{subfigure}{0.8\columnwidth}
\centering

\includegraphics[width=\maxwidth]{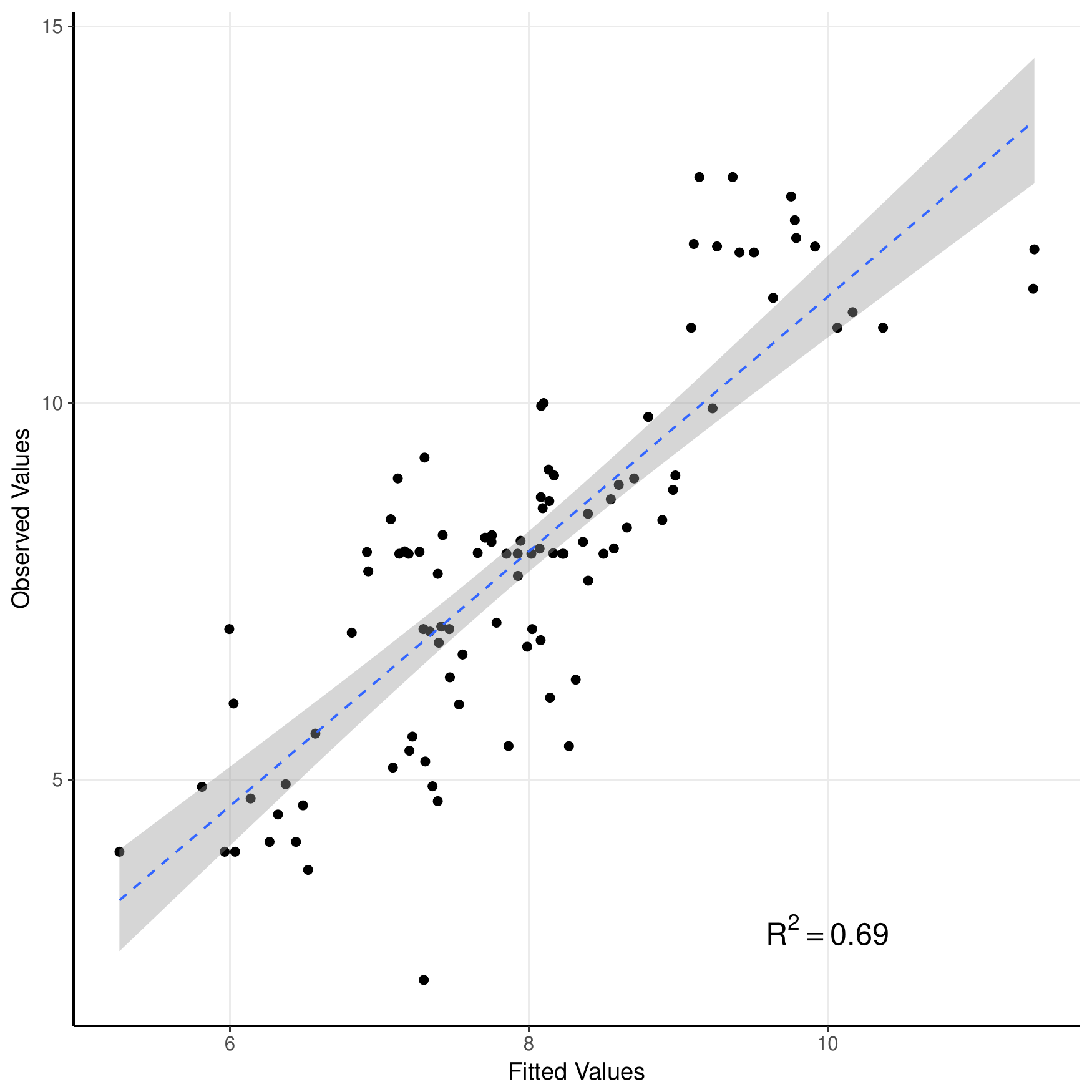} 
\end{subfigure}\caption{Fit of the stress--mental demand interaction model.}
\label{fig:assumFitLMEplain}
\end{figure}
}

\newcommand{\assumLMEReuse}{
\begin{figure*}[htb]

\includegraphics[width=\maxwidth]{figure/assumptions_lme_model_stress_reuse_interaction-1} 
\caption{Overview of LME assumption plots.}
\label{fig:assumLMEReuse}
\end{figure*}
}
\newcommand{\assumLMEReuseHist}{
\begin{figure}[htb]

\includegraphics[width=\maxwidth]{figure/assumptions_lme_model_stress_reuse_interaction-2} 
\caption{Histogram of Pearson residuals.}
\label{fig:assumLMEReuseHist}
\end{figure}
}
\newcommand{\assumLMEReuseQQplot}{
\begin{figure}[htb]

\includegraphics[width=\maxwidth]{figure/assumptions_lme_model_stress_reuse_interaction-3} 
\caption{QQPlot of Pearson residuals.}
\label{fig:assumLMEResueQQplot}
\end{figure}
}
\newcommand{\assumLMEReuseScatter}{
\begin{figure}[htb]

\includegraphics[width=\maxwidth]{figure/assumptions_lme_model_stress_reuse_interaction-4} 
\caption{Scatterplot of fitted values vs. residuals.}
\label{fig:assumLMEReuseScatter}
\end{figure}
}
\newcommand{\assumFitLMEReuse}{
\begin{figure}[htb]

\includegraphics[width=\maxwidth]{figure/assumptions_lme_model_stress_reuse_interaction-5} 
\caption{Comparison of fitted and observed values.}
\label{fig:assumFitLMEReuse}
\end{figure}
}
\newcommand{\assumFitLMEReusePlain}{
\begin{figure}[htb]
\centering
\begin{subfigure}{0.8\columnwidth}
\centering

\includegraphics[width=\maxwidth]{figure/assumptions_lme_model_stress_reuse_interaction-6} 
\end{subfigure}\caption{Fit of the stress--mental demand interaction model.}
\label{fig:assumFitLMEReusePlain}
\end{figure}
}



\newcommand{\esStress}{
\begin{figure}[htb]
\vspace{-2.5cm}
\includegraphics[width=\maxwidth]{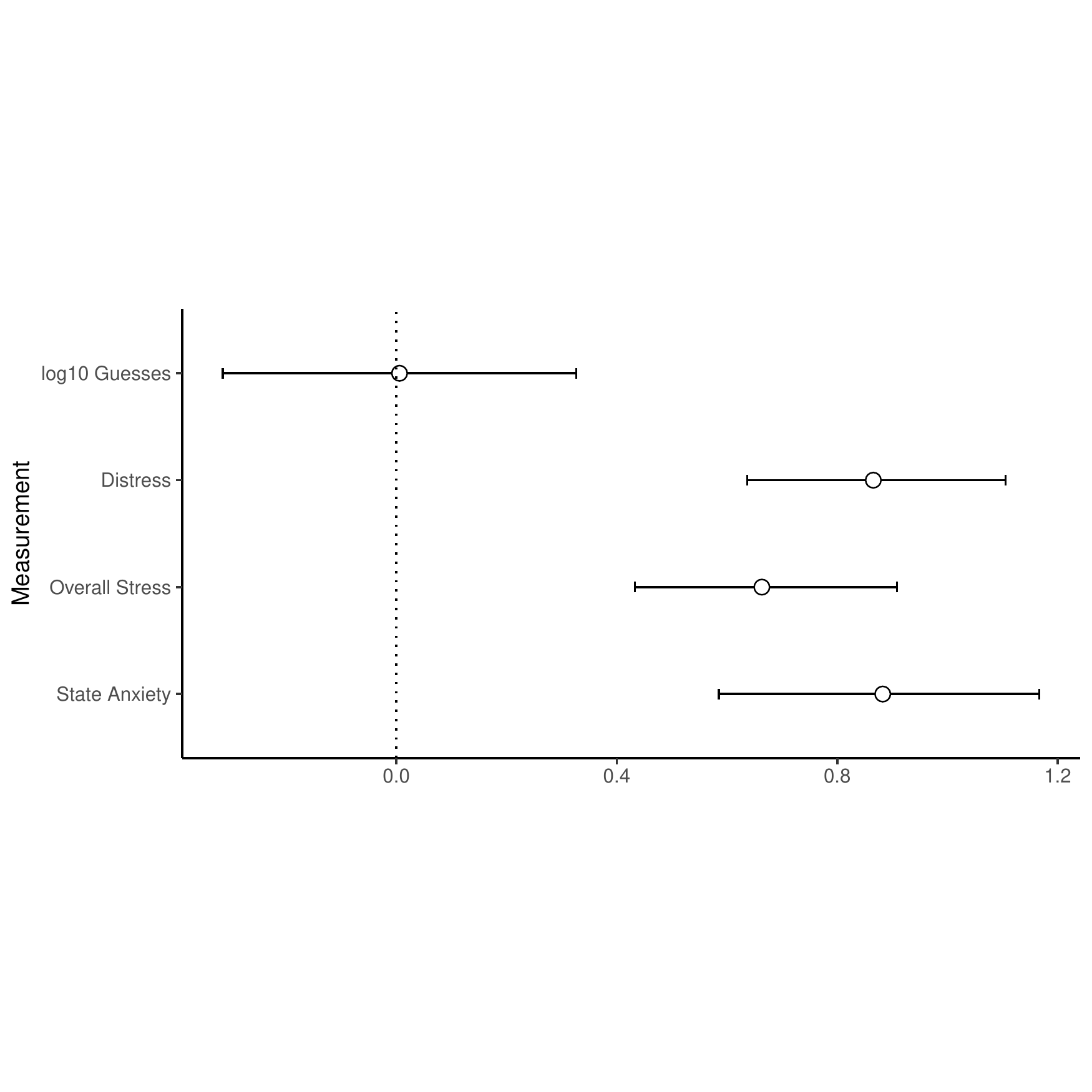} 
\vspace{-2.5cm}\caption{Comparison of the effects of the manipulations vis-`a-vis of the effect of stress on password choice. (Hedges' $g_\const{av}$, 95\% Confidence Intervals).}
\label{fig:esStress}
\end{figure}
}

%
%

\newcommand{\corrgramStress}{
\begin{figure}[htb]

\includegraphics[width=\maxwidth]{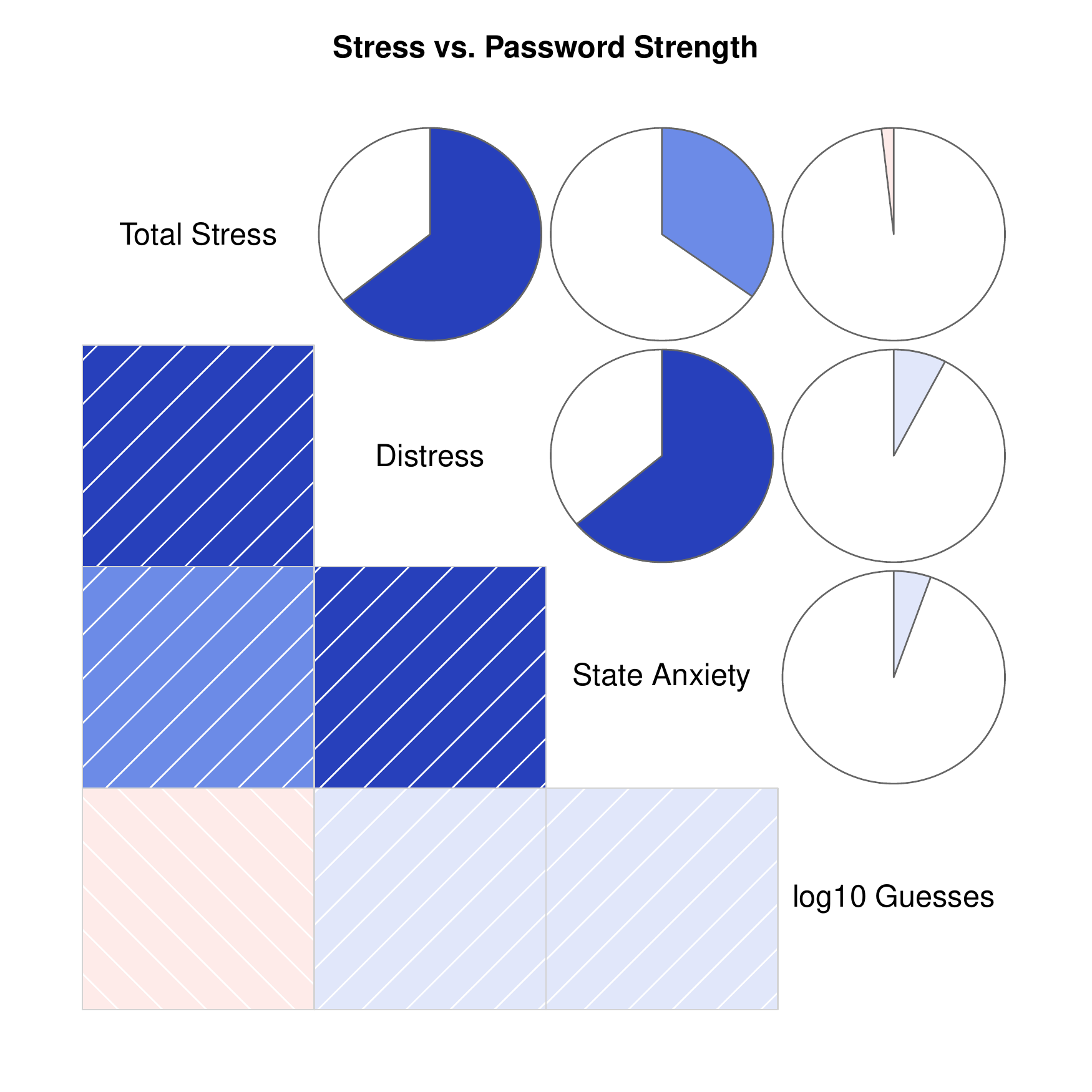} 
\caption{Corrgram of the variables irrespective of conditions.}
\label{fig:corrgramStress}
\end{figure}
}

%
%

\newcommand{\interactConditionReuse}{
\begin{figure}[htb]

\includegraphics[width=\maxwidth]{figure/debriefing-1} 
\caption{Interaction plot on Condition and Password Reuse.}
\label{fig:interactConditionReuse}
\end{figure}
}








\newcommand{\esComparisonLaser}{
\begin{figure}[htb]
\vspace{-2.5cm}
\includegraphics[width=\maxwidth]{figure/forest_plot-depletion-1} 
\vspace{-2.5cm}\caption{Comparision of previously reported depletion effects and the new studies of fear and stress (Hedges' $g$, 95\% Confidence Intervals).}
\label{fig:esComparisonLaser}
\end{figure}
}

\newcommand{\esComparisonAllCombined}{
\begin{figure*}[htb]
\begin{subfigure}{0.49\textwidth}
\vspace{-2.5cm}
\includegraphics[width=\maxwidth]{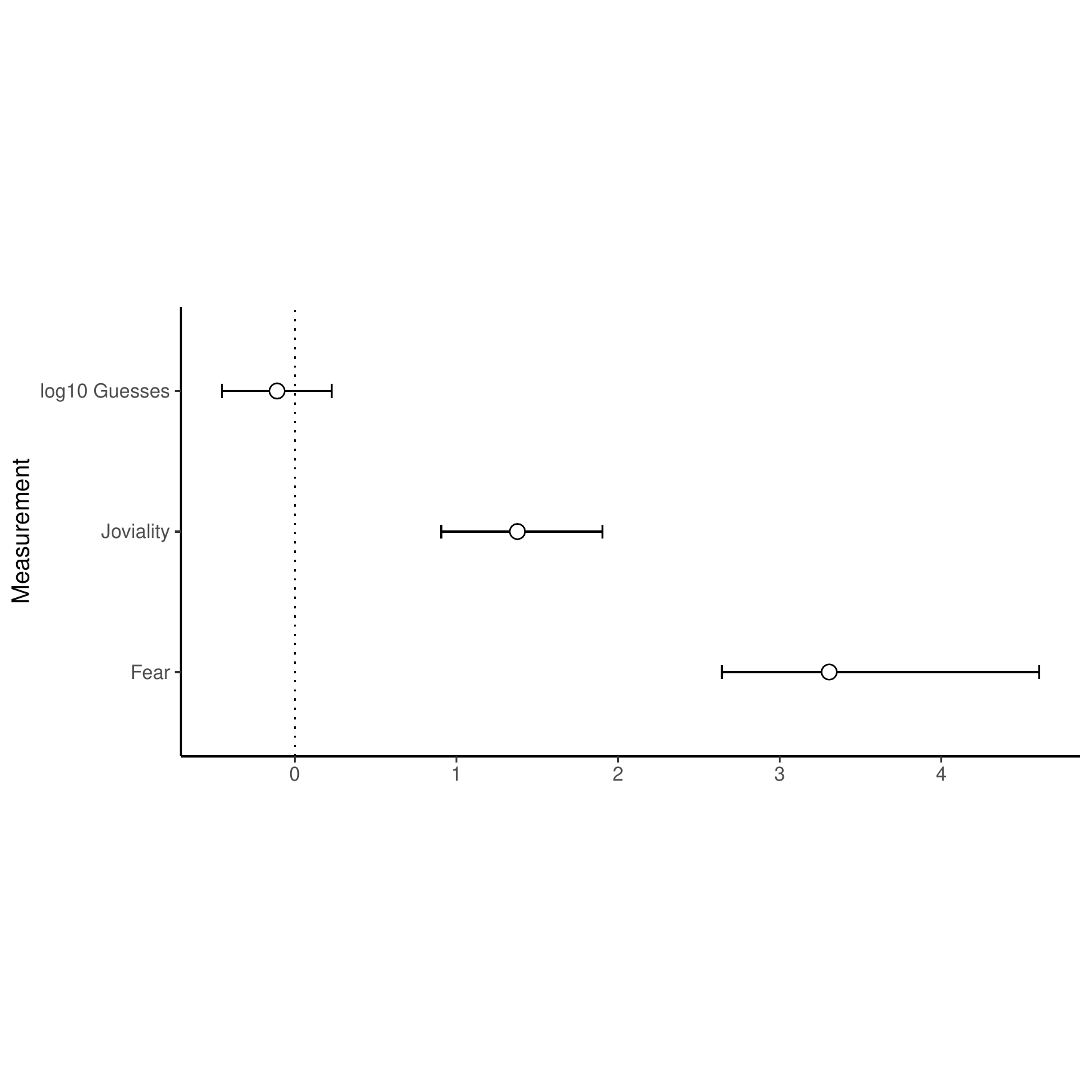} 
\vspace{-2.5cm}\subcaption{Study 1: Fear}
\label{fig:esFear}
\end{subfigure}~
\begin{subfigure}{0.49\textwidth}
\vspace{-2.5cm}

\includegraphics[width=\maxwidth]{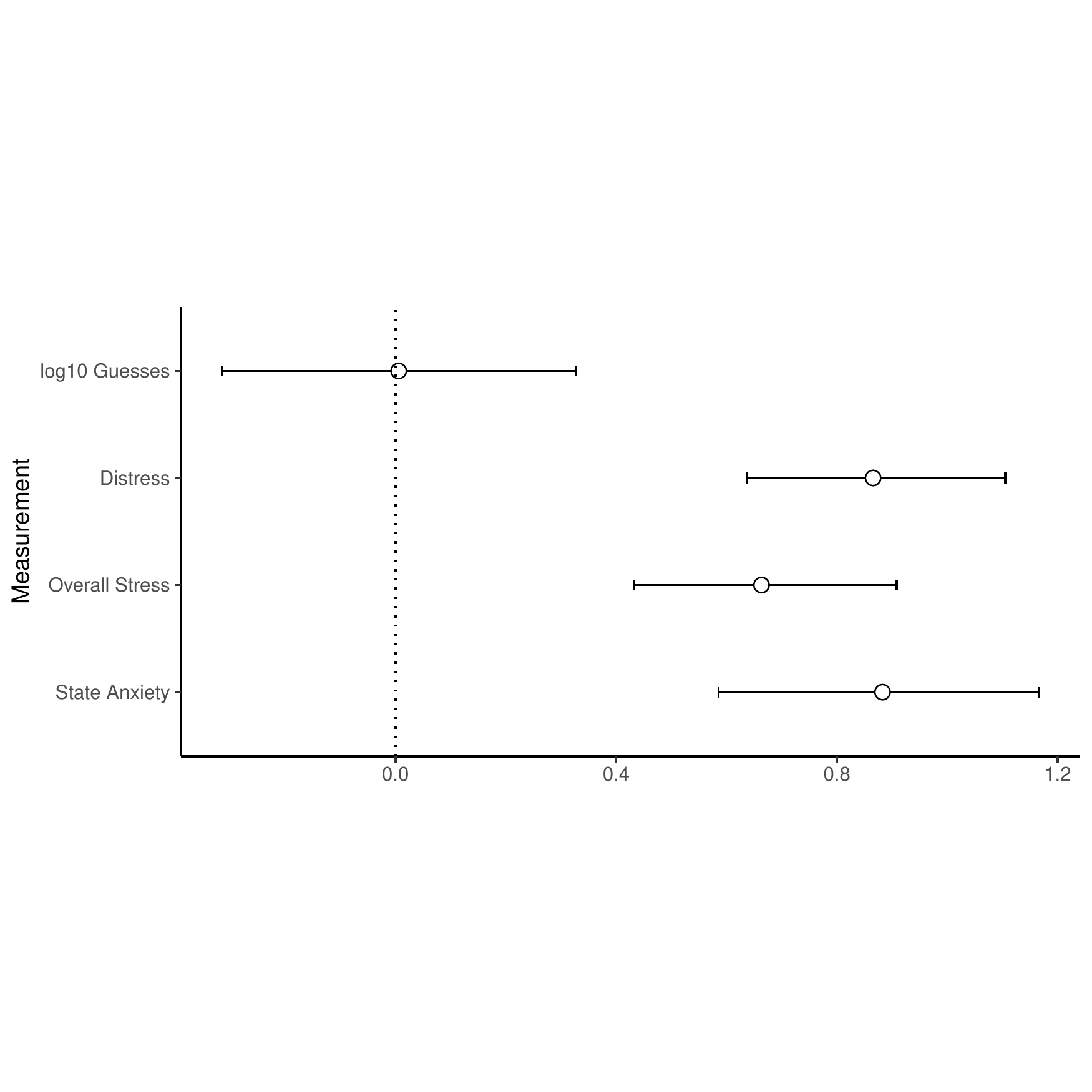} 
\vspace{-2.5cm}
\subcaption{Study 2: Stress}\label{fig:esStress}
\end{subfigure}
\begin{subfigure}{\textwidth}
\vspace{-2.5cm}

\includegraphics[width=\maxwidth]{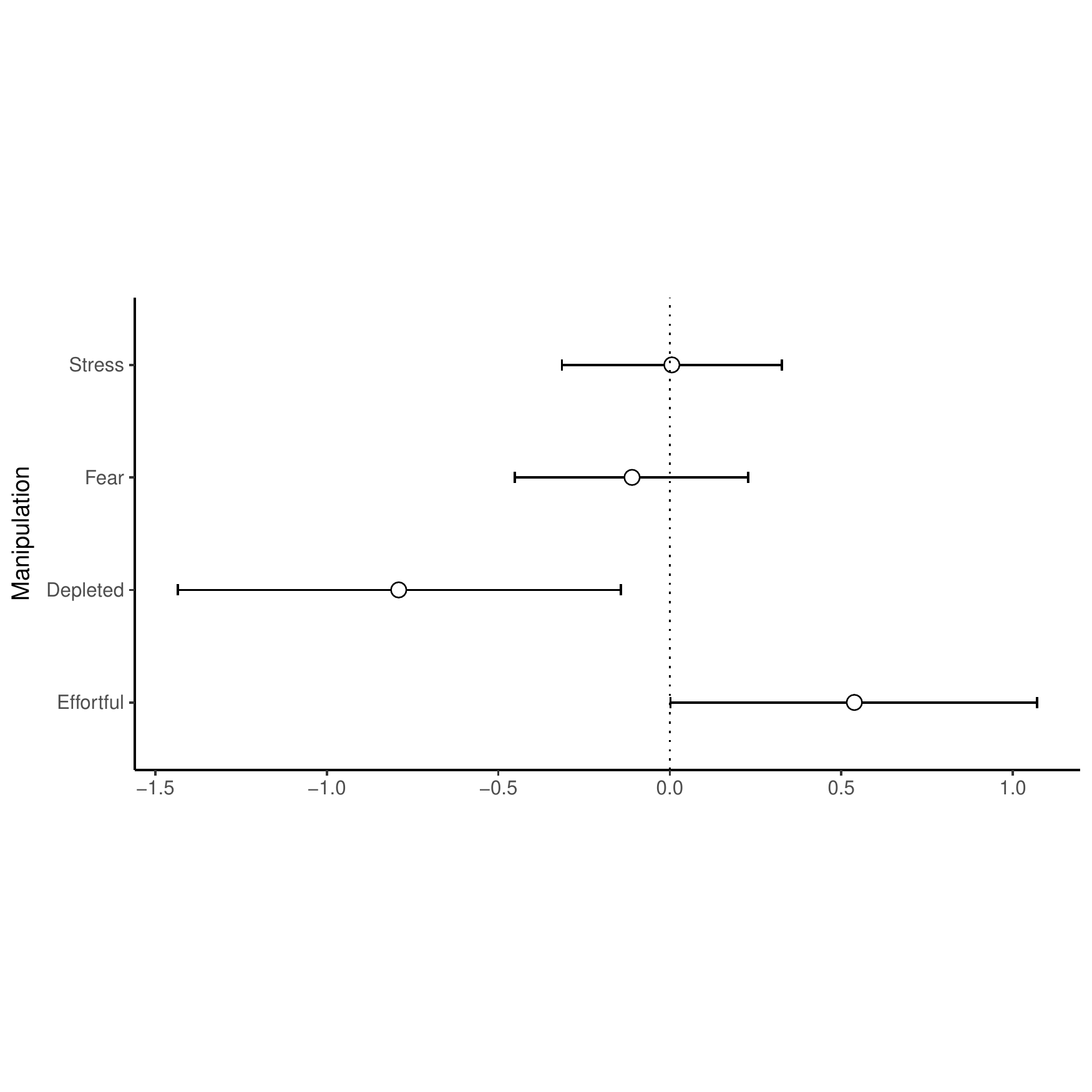} 
\vspace{-2.5cm}
\subcaption{Comparision of previously reported depletion effects.}
\label{fig:esComparisonLASER}
\end{subfigure}
\end{figure*}}

\newcommand{\networkMetaAnalysis}{
\begin{figure*}[htb]
\begin{subfigure}{0.70\textwidth}
\vspace{-4.5cm}
\includegraphics[width=\maxwidth]{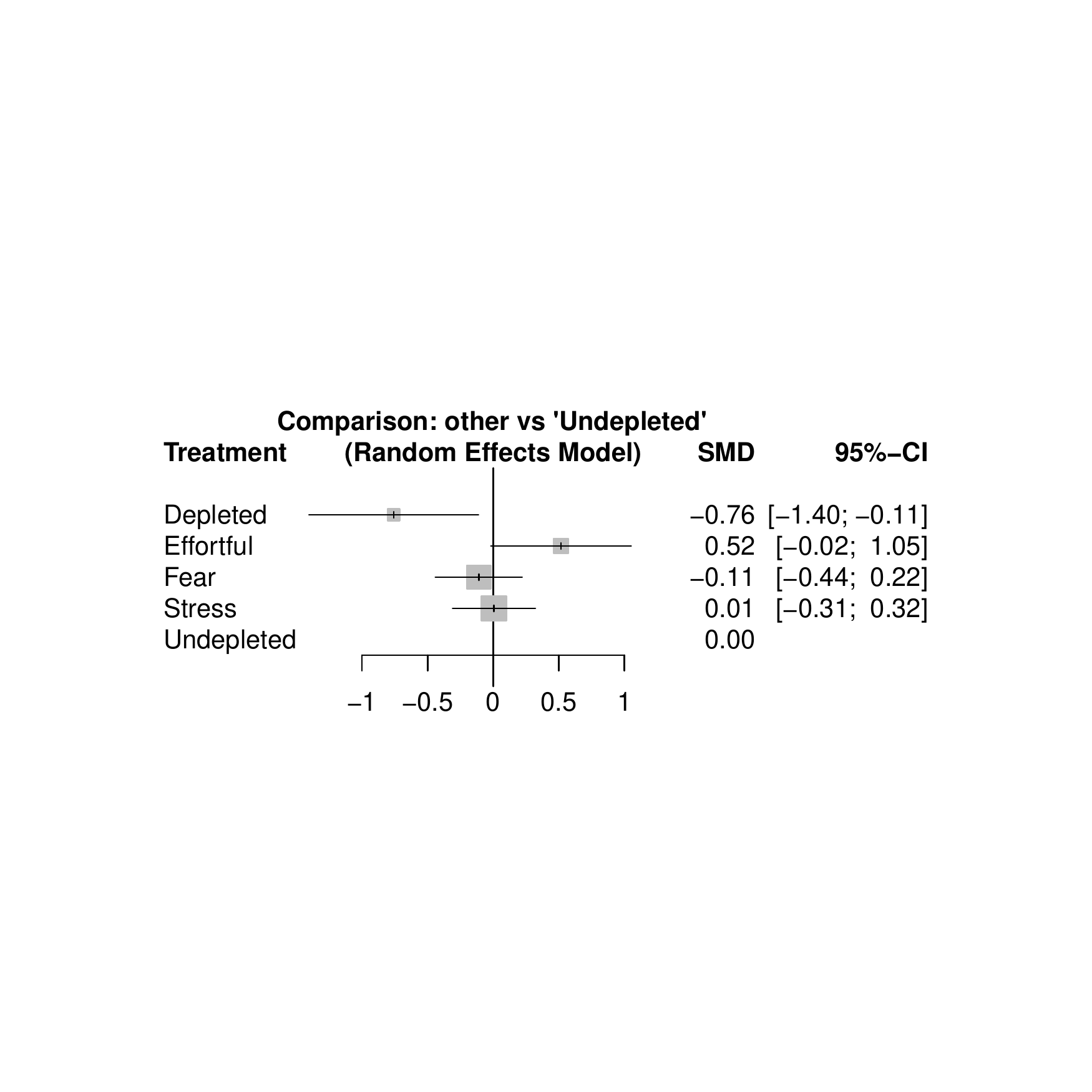} 
\vspace{-4.5cm}\subcaption{Forest Plot of Treatment Effects}
\label{fig:nma_forest}
\end{subfigure}~
\begin{subfigure}{0.29\textwidth}
\vspace{-2.0cm}

\includegraphics[width=\maxwidth]{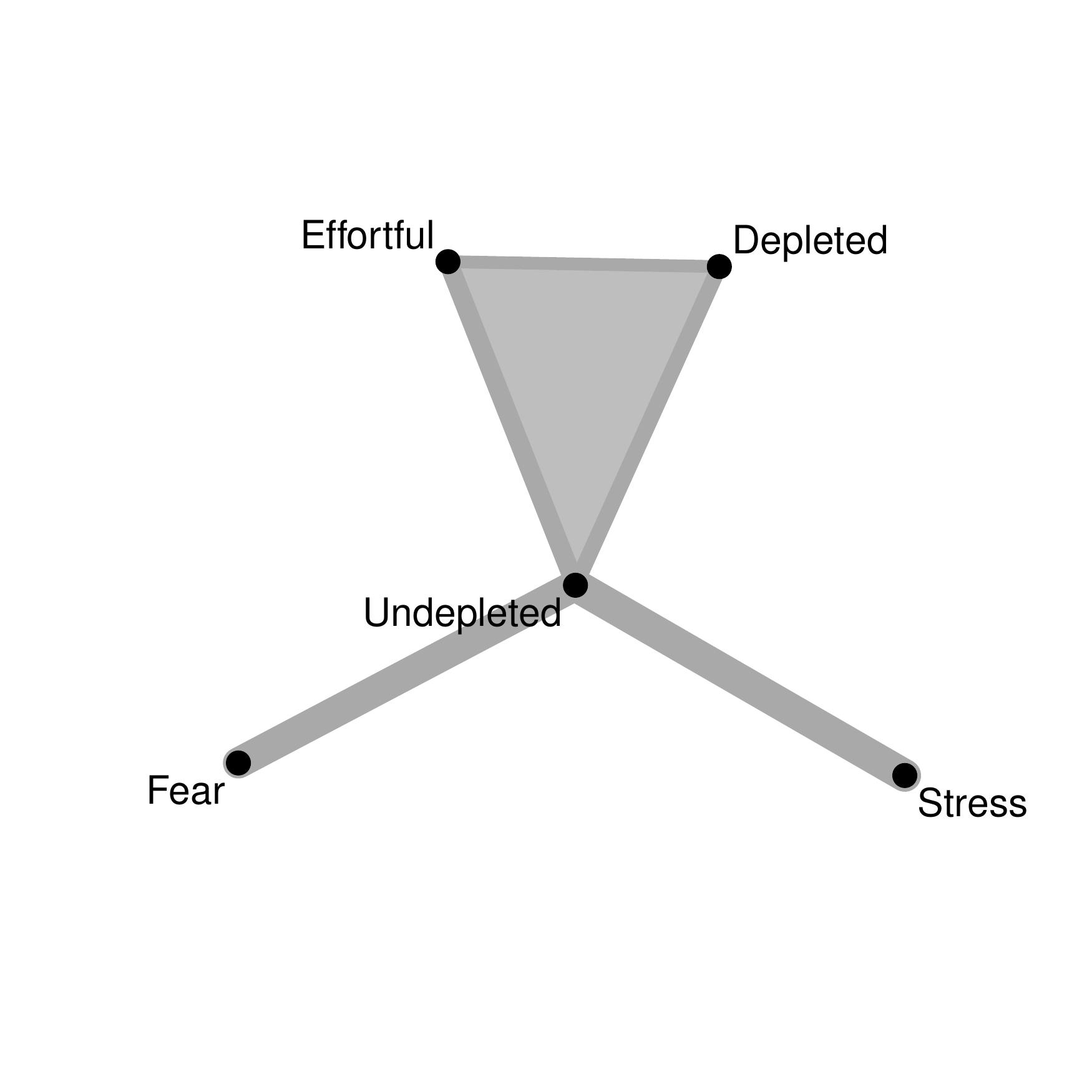} 
\vspace{-1.5cm}\subcaption{Network Graph}
\label{fig:nma_network}
\end{subfigure}
\caption{Network Meta Analysis of LASER'2016~\cite{GrCoAl2016} as well as this paper's Study 1: Fear and Study 2: Stress.}
\label{fig:networkMetaAnalysis}
\end{figure*}
}

\newcommand{\metaAnalysisOrder}{
\begin{figure}[htb]
\vspace{-2.7cm}
\includegraphics[width=\maxwidth]{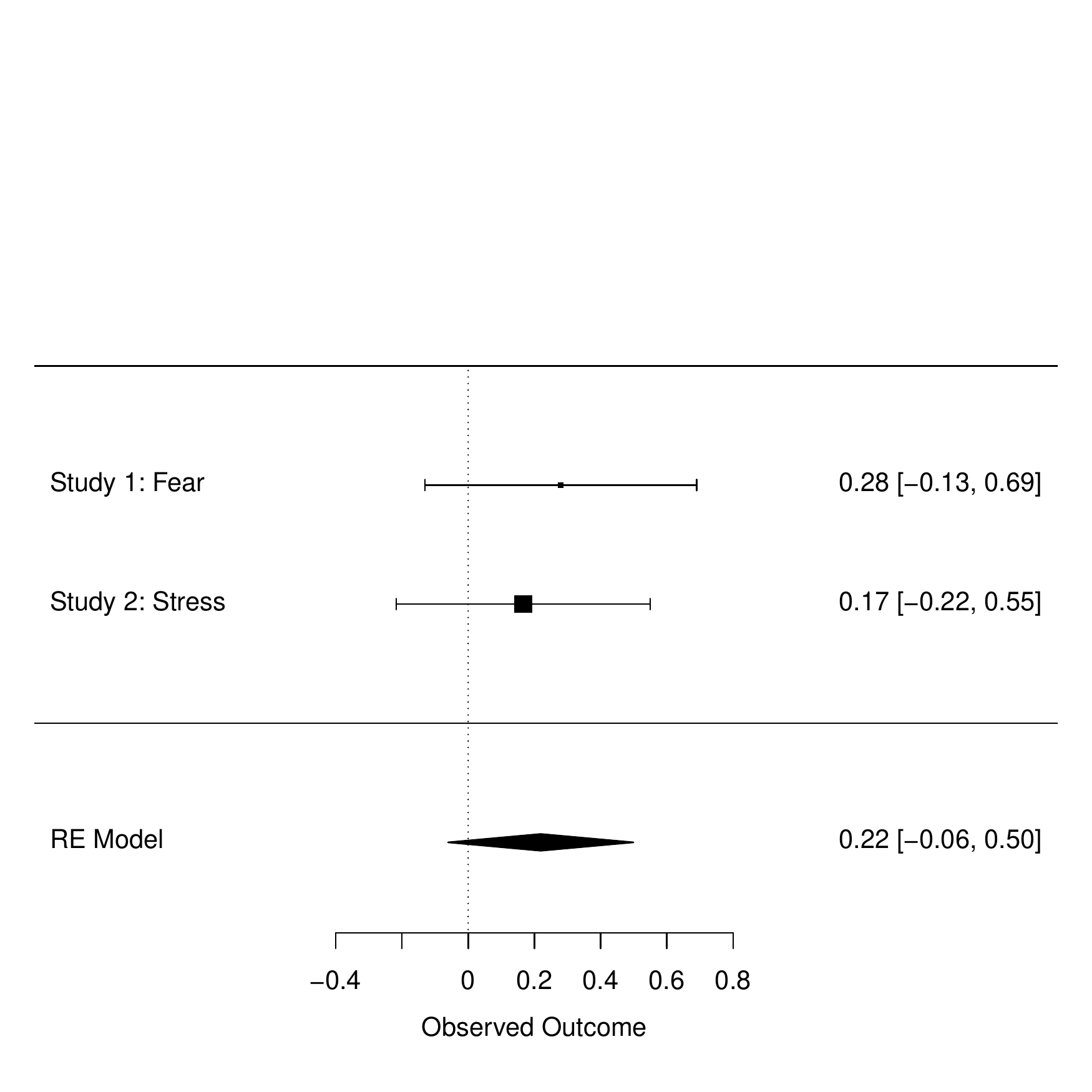} 
\vspace{-0.5cm}\caption{Forest Plot of order effects}
\label{fig:metaAnalysisOrder}
\end{figure}~
}

\date{}

\begin{DocumentVersionSTAST}
\title{Investigation of the Effect of Fear and Stress on Password Choice}
\end{DocumentVersionSTAST}
\begin{DocumentVersionTR}
\title{Investigation of the Effect of Fear and Stress on Password Choice (Extended Version)\thanks{Open Science Framework: \protect\url{https://osf.io/3cd9h}. This is the author's technical-report copy of this work. The definitive version has been published in Proceedings of the 7\textsuperscript{th} Workshop on Socio-Technical Aspects in Security and Trust (STAST'17), ACM Press, December 2018, pp. 3--15, \protect\url{https://doi.org/10.1145/3167996.3168000}.}}
\end{DocumentVersionTR}

\begin{DocumentVersionSTAST}
\author{Tom Fordyce}
\affiliation{%
  \institution{Newcastle University}
  \city{Newcastle upon Tyne}
  \country{UK}
}
\author{Sam Green}
\affiliation{%
  \institution{Newcastle University}
  \city{Newcastle upon Tyne}
  \country{UK}
}
\author{Thomas Gro{\ss}}
\affiliation{%
  \institution{Newcastle University}
  \city{Newcastle upon Tyne}
  \country{UK}
}
\end{DocumentVersionSTAST}

\begin{DocumentVersionTR}
\author{Tom Fordyce\\
School of Computing\\
Newcastle University, UK
\and Sam Green\\
School of Computing\\
Newcastle University, UK
\and Thomas Gro{\ss}\\
School of Computing\\
Newcastle University, UK}
\end{DocumentVersionTR}

\maketitle

\begin{abstract}
\noindent\textbf{Background.} 
The current cognitive state, such as cognitive effort and depletion~\cite{GrCoAl2016}, incidental affect or stress may impact the strength of a chosen password unconsciously.

\noindent\textbf{Aim.} 
  We investigate the effect of incidental fear and stress on the measured strength of a chosen password.

\noindent\textbf{Method.}
We conducted two experiments with within-subject designs measuring the \Zxcvbn~\cite{wheeler2016zxcvbn} \const{log10} number of guesses as strength of chosen passwords as dependent variable. In both experiments, participants were signed up to a site holding their personal data and, for the second run a day later, asked under a security incident pretext to change their password.
\begin{inparaenum}[(a)] 
\item \textbf{Fear.} $N_\const{F} = 34$ participants were exposed to standardized fear and happiness stimulus videos in random order. 
\item \textbf{Stress.} $N_\const{S} = 50$ participants were either exposed to a battery of standard stress tasks or left in a control condition in random order. 
The \Zxcvbn password strength was compared across conditions. 
\end{inparaenum}

\noindent\textbf{Results.}
We did not observe a statistically significant difference in mean \Zxcvbn password strengths on fear (Hedges' $g_{\const{av}} = -0.11$, 95\% CI $[-0.45, 0.23]$) or stress (and control group, Hedges' $g_{\const{av}} = 0.01$, 95\% CI $[-0.31, 0.33]$). However, we found a statistically significant cross-over interaction of stress and TLX mental demand.
  
\noindent\textbf{Conclusions.}
While having observed negligible main effect size estimates for incidental fear and stress, we offer evidence towards the interaction between stress and cognitive effort that vouches for further investigation.
\end{abstract}

\section{Introduction}
While recently heralded as its days being numbered~\cite{bonneau2012quest,bonneau2010password}, username and password are still the predominant authentication mechanism. At the same time, password choice makes for an archetypical security task, in which users are focused on reaching a primary goal (accessing a Web site), while the security of the password is only a secondary goal. Hence, results on users' password choice can be informative beyond the act of choosing a password itself.

While there is a body of research on user habits in password choice~\cite{florencio2007large}, recent research also aimed at investigating how the user's current cognitive or affective state impacts the strength of chosen passwords. Imagine a user being depleted after a long day's work or being stressed out after a painful social interaction.

Given that such cognitive and affective states are found to impact \emph{inter alia} executive function, working and declarative memory, effects on security decision making are plausible. This may hold for for security specialists fighting a security incident, experiencing stress and cognitive depletion in the process, as much as for off-the-street users, experiencing stress and cognitive depletion in everyday life. The current state of the user could then unconsciously impair the strength of the password choice.

While Flor{\^e}ncio et al.~\cite{florencio2014password} already considered password strategies for finite-effort users, Gro{\ss} et al.~\cite{GrCoAl2016} investigated the effect of cognitive effort and depletion~\cite{Kah1973,BauBra1998} on password choice in depth, concluding that cognitive effort is a necessary condition for strong passwords. By their account, cognitive depletion would impair password choice significantly. 

We investigate the \emph{effect of fear and stress on password choice} as an archetypical security task. We focus on incidental stress, that is, stress that is unrelated to the security task at hand. 

It is rather elusive to induce stress independently from cognitive effort, because the host of stress induction instruments cause require the participant to exert cognitive effort at the same time. The reason for that is that by Baumeister's \emph{limited strength model}~\cite{BauBra1998} also governs the use of willpower. Stress induction techniques, however, no matter whether they are cognitive~\cite{li2015laboratory}, physical~\cite{matthews1988influences}, or social~\cite{kirschbaum1993trier} require the participant to exert willpower as well to keep going, to persevere in the task.

For that reason, we have not only designed an experiment that induces stress, but also a second experiment that induces fear. We have chosen to induce fear with a standardized video stimulus, which does not yield any cognitive effort whatsoever.

In this paper, we contribute two studies establishing the effect of incidental fear as well as of stress on password choice. To our knowledge, these are the first studies that induce fear and stress in a password choice scenario. In addition, we investigate the interaction between stress and cognitive effort. We offer research synthesis between these two studies and the earlier work on cognitive effort in a network meta analysis to put these results in context of one another.

\section{Preliminaries}
This study is founded on the principles of parameter and interval estimation~\cite{cumming2013understanding}, that is, we are less interested in null-hypothesis significance testing (NHST). 
Instead, we seek to quantify the magnitude of effects and to offer confidence intervals to bracket the likely effect sizes in the population.

While NHST has received its share of criticism and observed fallacies~\cite{Nickerson2000null,gardner1986confidence}, we aim at gaining robustness through complementing NHST with estimation methods~\cite{cumming2013understanding}.

\begin{DocumentVersionTR}
\subsection{Parameter Estimation}
We use standardized effect size metrics to ascertain the magnitude of observed effect.
Such an effect size estimates the difference between means in the population standardized by the observed variance. A prevalent recommendation in the community~\cite{cumming2013understanding,lakens2013calculating} for correlated (within-subjects) means is Hedges' $g_\const{av}$:

\begin{align*}
g_\const{av} & = \left( 1 - \frac{3}{4(\vari{df}) - 1}\right) \left(\frac{M_\const{diff}}{\frac{\vari{SD}_1 + \vari{SD}_2}{2}}\right)
\end{align*}

\subsection{Interval Estimation}
We further seek to ascertain the confidence interval of the effect size. We consider 95\% confidence intervals, which yield that if the study were run infinitely often, the true population effect size will be captured by the interval 95\% of the runs.

For our correlated samples case, we use a recommended estimation method by Algina and Keselman~\cite{algina2003approximate,cumming2013understanding} to gain an interval estimate on the population effect size.
\begin{inparaenum}[(a)]
\item We estimate the non-centrality parameter $\lambda$ with a $t$ distribution,
\item compute the confidence interval on the $\lambda$ and
\item then finally multiply its confidence limits with the following factor to obtain the final confidence interval on the effect size:
\[ \sqrt{\frac{2(\vari{SD}_1^2 + \vari{SD}_2^2 - \vari{COV}_{12})}{n(\vari{SD}_1^2 + \vari{SD}_2^2)}}. \]
\end{inparaenum}
\end{DocumentVersionTR}

\begin{DocumentVersionSTAST}
\subsection{Effect Size and Interval Estimation}
We use effect sizes for correlated (within-subjects) means with Hedges' $g_\const{av}$. Therein, we follow recommendations of Cumming~\cite{cumming2013understanding} and Lakens~\cite{lakens2013calculating}. We use an interval estimation for correlated samples by Algina and Keselman~\cite{algina2003approximate,cumming2013understanding}.

The extended technical report for this paper~\cite{FoGrGr2017a_TR} contains the exact formula and reasoning for these methods.
\end{DocumentVersionSTAST}

\subsection{Network Meta Analysis}
Meta analysis refers to a set of statistical techniques to combine the results of multiple studies~\cite{cumming2013understanding,cooper2009handbook}.
\emph{Network meta analysis} (NMA), in particular, specializes on combining results of studies with different treatments in the same meta analysis simultaneously.
For instance, in this paper we consider studies which all studied the impact of certain ``treatments,'' such as stress, fear or cognitive effort, on password strength.
Schwarzer et al.~\cite{Schwarzer2015_NMA} offer an introduction to the method itself, while Binod et al.~\cite{Binod2014} offer a good overview of such techniques available in \textsf{R}. In this work, we use the \textsf{R} package \textsf{netmeta}, which implements frequentist and graph-theoretical techniques following R{\"u}cker~\cite{rucker2012network}.

\section{Background}

\subsection{Password Choice}
In terms of overall user password behavior, the average user is said to have 6.5 passwords, each shared across 3.9 different sites, each user has 25 accounts requiring passwords and type 8 passwords per day \cite{florencio2007large}. 

It has been argued that the recall in password authentication itself is a humanly impossibly task, because non-meaningful items are inherently difficult to remember~\cite{sasse2001transforming}. This is aggravated by typical password policies asking users to comply to arcane procedures, such as monthly password reset. Such policies cause users to feel frustrated and security-fatigued~\cite{hoonakker2009password_brief}. Consequently, users are naturally employing alternative strategies such as writing passwords down, incrementing the number in the password at each reset \cite{adams1999users}, storing passwords in electronic files and reusing or recycling old passwords~\cite{hoonakker2009password_brief}. 

Password reuse, in particular, having been observed as overall tendency and individual preference, has received further attention in recent research~\cite{das2014tangled}.

While it is possible to create strong and meaningful passwords using pseudo-random combinations of letters, numbers and characters that are meaningful only to the owner~\cite{zviran1993comparison}, from qualitative research it was found that four to five passwords are the most a typical user can be expected to use effectively~\cite{adams1999users}.

\subsubsection{Password Strength Measurement}
There are controversies around how to measure password strength soundly and reliably~\cite{bonneau2012science}, where earlier methods such as the NIST password entropy heuristic (NIST Special Publication 800-63~\cite{NIST2004}) have received criticism.

A recent research direction is to practicably estimate the number of guesses needed by an adversary to break the password~\cite{kelley2012guess,wheeler2016zxcvbn}.

In this work, we considered the Password Guessability Service (PGS)~\cite{kelley2012guess} hosted by CMU as well as the Dropbox password meter \Zxcvbn~\cite{wheeler2016zxcvbn}, which both output an estimate of the number of guesses. We chose \Zxcvbn as final measurement instrument.

\subsection{Affect}
Affect is the experience of an emotion or feeling, where we also consider the observable affect, that is, behavior serving as indicator of affect.

While there exist a number of conceptualizations for affect, mood and emotions as well as a body of mature research in psychology~\cite{davidson2009handbook,lewis2010handbook}, we consider the Russel's core affect conceptualization~\cite{russell2003core} as a guiding work.
Russel considers the dimensions activation-deactivation and pleasure-displeasure as foundational.

We make a distinction between two kinds of affect~\cite{peters2006affect}:
\begin{compactitem}
\item \emph{integral affect} (also task-related affect) refers to the experienced feelings with respect to a stimulus, and 
\item \emph{incidental affect} refers to feelings such as
mood states that are independent of a stimulus but can be misattributed to it or can influence
decision processes.
\end{compactitem}

For our enquiry into fear and stress, Russel's core affect model is especially interesting as it allows us to classify distress and fear in the common quadrant of displeasure/activation.

\subsection{Fear and Fear Appeals}
Affects have been considered in security mostly concerning fear, especially in relation to fear appeals. \emph{Fear appeals}~\cite{rogers1983cognitive,floyd2000meta,witte1992putting} are messages designed to motivate a certain behavior by eliciting fear. 

While fear appeals are are operating with integral/task-related fear, we consider incidental fear and stress in this work.

Boss et al.~\cite{boss2015users} gave an overview of fear appeals research in information security, pointing out that most prior studies in the space did not measure fear directly. In general, Ruiter et al.~\cite{ruiter2014sixty} pointed out that coping information is bound to be more important in yielding a protection motivation than risk warnings and fear arousal.

\subsubsection{Affect Elicitation}
There has been considerable research on affect elicitation. For instance, the first ten chapters of the \emph{Handbook of Emotion Elicitation and Assessment}~\cite{coan2007handbook} are concerned with different elicitation methods. Not all elicitation methods are created equal, though. Westermann et al. found considerable differences in effectiveness and validity of mood induction procedures (MIP)~\cite{westermann1996relative}. In this study, we focus on Film MIP without instructions which was a category recommended by Westermann et al. From inspecting this past research, we conclude that the chosen MIPs are sound to elicit affect in participants.

Affect elicitation with stimulus films has received a systematic treatment in affective psychology~\cite{gross1995emotion}, where our stimulus videos are drawn from Rottenberg et al.'s work~\cite{ray2007emotion}. Here, we find particular recommendations for video segments validated for designated target emotions, such as fear or happiness. 

We observe that it was reported among others by Rottenberg et al.~\cite[pp. 103]{ray2007emotion} that fear is a challenging emotion to elicit discretely. In their experiments with stimulus films, such as \emph{The Silence of the Lambs}, they found that the films also elicited tension and interest in equal measures, however they argued that the confluence of fear, interest and tension is indeed natural.

We note that affect elicitation with stimuli films has been validated in within-subject experiment designs~\cite{ray2007emotion}. The variable ``within-subject design'' was taken into account in a comprehensive meta analysis on MIPs~\cite{westermann1996relative}, where the impact of this variable was not found to be statistically significant ($r = .08$ for films/story MIPs.).

\subsubsection{Affect Measurement}
While Coan and Allen~\cite{coan2007handbook} give an overview of a range of affect measurement methods, we focus on self-report instruments such as the Positive and Negative Affect Schedule (PANAS-X)~\cite{watson1999panas}. The questionnaire primarily measures the valence of reported affect (positive or negative) and its content (e.g., fear or joviality). The questionnaire was designed to reliably measure affect while still being easily administered. The full 60-item schedule usually takes participants about 10 minutes.

\subsection{Stress}
Selye~\cite{selye1974stress} originally defined stress as the \emph{the nonspecific response of the body to any demand made upon it}, and distinguished between eustress and distress, hyperstress and hypostress.

It was postulated that some levels of stress may improve performance and that performance will deteriorate once an optimal range of arousal is passed. With respect to arousal in habit formation, such a relationship was also formulated as the Yerkes-Dodson law~\cite{teigen1994yerkes}.

\subsubsection{Stress Elicitation}
Stress can be elicited in experiments predominantly by cognitive, physical or social instruments. Liao and Carey~\cite{li2015laboratory} offer an overview of lab stress elicitation methods.

Existing experimental protocols often combine validated cognitive, physical and social stress tasks, where such tasks have been used to induce psychological stress and receive a cardiovascular response~\cite{li2015laboratory}. In addition, there exist instruments, such as \emph{Trier Social Stress Test} (TSST)~\cite{kirschbaum1993trier}, which are elaborate protocols to induce stress in multiple stages.

\subsubsection{Stress Measurement}
While stress can be measured psychophysiologically from heart rate variability and skin conductance, a number of instruments have been proposed to measure stress in self-report questionnaires.

Partially, researchers used affective instruments, such as the State-Trait Anxiety Inventory (STAI)~\cite{spielberger1970manual} to gauge stress.

Partially, researchers developed and used specialized stress scales, such as the Dundee Stress State Questionnaire (DSSQ)~\cite{matthews1999validation} or its short form, the Short Stress State Questionnaire (SSSQ)~\cite{helton2004validation,helton2015short}.

\subsection{Cognitive Effort}
Research in cognitive effort has a long history in psychology, for instance starting from Kahneman's work on effort and attention~\cite{Kah1973}.

Baumeister et al.~\cite{BauBra1998} first proposed that human beings have a limited store of cognitive energy, formulated in the \emph{limited strength model of cognitive effort}. 
Willpower and self-control are said to draw from this inner resource as well as cognitively hard tasks.
Examples include controlling attention, emotions, impulses, thoughts and cognitive processing, choice and volition and social processing~\cite{baumeister2007strength}. In general, all tasks that are cognitively effortful draw from the limited cognitive energy.

\subsubsection{Cognitive Effort Measurement}
Cognitive effort as a construct can, for instance, be measured by tasks that require cognitive effort themselves. In those examples the error rate on the tasks will indicate cognitive depletion. These measurement methods come at the disadvantage that they change the participants' state by inducing cognitive depletion themselves.

Baumeister et al. consequently proposed to use a Brief Mood Introspection Scale (BMIS)~\cite{mayer1988experience,BauBra1998} or a short form~\cite{tice2007restoring} as a proxy to measure cognitive effort through items such as ``being tired'' and ``being worn-out.''

Other research focused on the measurement of the cognitive effort needed for a task, the \emph{task load}. A notable measurement instrument along this lines that stood the test of time is the NASA Task Load Index (TLX).

\section{Aims}
\begin{researchquestion}[Study 1: Fear]
To what extent do elicited affects happiness and fear impact password strength?
\end{researchquestion}
Table~\ref{tab:ops_fear} gives an overview of the operationalization of this research question.
As independent variable (IV), we have elicited affect with the two levels: Fear and Happiness.

We intend to check that the manipulation was successful by evaluating the Positive and Negative Affect Schedule (PANAS-X)~\cite{watson1999panas} on \textsf{fear} and \textsf{joviality}.
The null hypothesis of this manipulation check is $\const{H}_{\const{mc,F,0}}$: \emph{There is no mean difference between either \textsf{fear} or \textsf{joviality} between conditions.}
We call the manipulation successful if this null hypothesis is rejected.

The null hypothesis of the overall experiment is $\const{H}_{\const{F,0}}$: \emph{There is no mean difference in \Zxcvbn \const{log10} guesses between conditions.}
The corresponding alternative hypothesis $\const{H}_{\const{F,1}}$ is \emph{The \Zxcvbn \const{log10} guesses differ between conditions.}

\begin{table*}[tb]
\centering
\footnotesize
\caption{Operationalization of Study 1: Effect of Fear on Password Strength.}
\label{tab:ops_fear}
\begin{tabular}{llll}
\toprule
                & Levels	    & Instrument & Intervention/Variable\\
\midrule
IV: Affect	& Fear 	    & Stimulus Video~\cite{ray2007emotion} & \emph{Silence of the Lambs} \\
		& Happiness & 								&\emph{When Harry Met Sally} \\
\midrule
IV Check & Fear		     & PANAS-X~\cite{watson1999panas}    & \textsf{fear} \\
	        & Happiness 	      & 						    & \textsf{joviality} \\
\midrule
DV: Pwd Strength &		      &\Zxcvbn~\cite{wheeler2016zxcvbn} & \const{log10} Guesses \\
\bottomrule
\end{tabular}
\end{table*}

\begin{researchquestion}[Study 2: Stress]
To what extent does elicited stress impact password strength?
\end{researchquestion}
Table~\ref{tab:ops_stress} operationalizes this research question.
As independent variable (IV), we have elicited stress with the two levels: Stress and Control.

We check the success of the manipulation with two kinds of instruments: the Short State Stress Questionnaire (SSSQ)~\cite{helton2015short}, considering overall \textsf{stress} and \textsf{distress}, as well as the State-Trait Anxiety Inventory (STAI)~\cite{spielberger1970manual}, considering \textsf{state\_anxiety}.

The null hypothesis of this manipulation check is $\const{H}_{\const{mc,S,0}}$: \emph{There is no mean difference between either \textsf{stress}, \textsf{distress}, or \textsf{state\_anxiety}.}
We call the manipulation successful if this null hypothesis is rejected.

The null hypothesis of the overall experiment is $\const{H}_{\const{S,0}}$: \emph{There is no mean difference in \Zxcvbn \const{log10} guesses between stressed and control condition.}
The corresponding alternative hypothesis $\const{H}_{\const{S,1}}$ is \emph{The \Zxcvbn \const{log10} guesses differ between conditions.}

\begin{table*}[tb]
\centering
\footnotesize
\caption{Operationalization of Study 2: Effect of Stress on Password Strength.}
\label{tab:ops_stress}
\begin{tabular}{llll}
\toprule
                & Levels	    & Instrument & Intervention/Variable\\
\midrule
IV: Stress	& Stress 	    & Serial Subtraction Task~\cite{li2015laboratory} & cont. sub. 7 from 9095, 1.5 min\\
		&		    &  & cont. sub. 13 from 5245, 1.5 min \\
		&		    & Isometric Handgrip Task~\cite{matthews1988influences}& 30\% max strength, 2.5 min \\
		&		    & Social Stress akin to TSST~\cite{kirschbaum1993trier} & results judged; fail: start over \\
		& Control & Balanced: Serial addition   & cont. add 7 to 9095, 1.5 min\\
		&		&					& cont. add. 13 to 5245, 1.5 min\\
		& 		& Balanced: Isometric Handgrip & 30\% max strength, till discomfort\\
\midrule
IV Check & Total Stress	& SSSQ~\cite{helton2004validation,helton2015short,matthews1999validation}
    & \textsf{stress} \\
	        & Distress 	      & 						    & \textsf{distress} \\
	        & Anxiety		     & STAI~\cite{spielberger1970manual} & \textsf{state\_anxiety} \\
\midrule
Cognitive Effort	        & Task Load & TLX~\cite{hart1988development,hart2006nasa} & \textsf{tlx}\\
	        & Mental Demand &  & \textsf{tlx\_mental}\\
\midrule
DV: Pwd Strength &		      &\Zxcvbn~\cite{wheeler2016zxcvbn} & \const{log10} Guesses \\
\bottomrule
\end{tabular}
\end{table*}

\section{Method}
For reproducibility and scientific integrity, the study has been registered at the Open Science Framework (OSF)\footnote{\url{osf.io/3cd9h}}. The comprehensive report~\cite{FoGrGr2017_AnaRep} of all analyses computed is registered at the OSF, as well. Analyses, graphs and statistical reporting in this paper were computed directly from the data using the \textsf{R} package \textsf{knitr}.

\subsection{Common Approach}
\subsubsection{Within-Subjects Lab Experiment}
Both Studies were conducted simultaneously as within-subject experiments with two conditions, that is, each participant goes through both conditions. We determined in a constrained random block assignment in which order each participant was exposed to the conditions, while maintaining balanced sub-sample sizes.

We stress that affect elicitation and mood induction procedures have been experimentally and meta-analytically validated for within-subject settings~\cite{ray2007emotion,westermann1996relative}. To ensure that a MIP stimulus of a preceding session does not confound a subsequent session, we leave a break of 24 hours between sessions.

We have further chosen to run the studies as a lab experiment and not as a study on Amazon Mechanical Turk (AMT), primarily because the stress manipulations required physical presence of the participants. Given that we induced stress as well as fear, we deemed it ethical that the experiments were conducted by a physically-present experimenter, who could offer information about the experiment in person, allow participants to withdraw from the experiment in dignity and ensure that participants are not leaving the experiment overly disturbed.

\subsubsection{Sampling}
The participants for both studies were recruited independently from one another.

We have chosen to run the experiments as within-subject design with the given sample sizes to gain at least $80\%$ power for medium effect sizes, after an \emph{a priori} power analysis. A two-tailed dependent-samples $t$-test requires a sample size of $N=34$ to reach $80\%$ power at a significance level of $\alpha = .05$. The effect size estimates were chosen to be smaller than observed medium-large effects reported by Gro{\ss} et al.~\cite{GrCoAl2016}.

In both studies, the participants were largely recruited from university staff and students.

\subsubsection{Overall Procedure}
In both studies, participants are asked to register on a Web site, which stores personal information as well as sensitive data about them, such as personality traits and psychometric test results. The participants are made aware of the sensitivity of the data. There was no password policy imposed on the participants. Participants were asked to ``choose'' a password, that is, they were not asked to refrain from reusing prior passwords. 

The experiments are conducted over two days, to let the effects of prior manipulations subside. The break between both runs was at least 24 hours, but no more than five days. All participants returned for the second run. No participant withdrew from the study.

When the participants returned for the second day, they are informed that they need to set a new password for their personal data under the pretext of a security incident. The system enforced that they could not repeat exactly the same password.

For both experiments, we first elicited an affective state, then asked the participants to register an account with a password, and finally had a post-task manipulation check to evaluate how well the affect elicitation worked. We note here that the manipulation check was deliberately placed \emph{after} the password task, to ensure that the affect during the task was at least as strong as the one measured.

The participants were only told in the debriefing of each study that the experiment's true purpose was about password strength. Study 2 included a debriefing questionnaire on the modalities of password choice made.

\subsubsection{Password Strength Measurement}
Password strength was measured as \const{log10} number of password guesses as evaluated with an offline \Zxcvbn~\cite{wheeler2016zxcvbn} with standard dictionaries. 

\subsection{Ethics}
Both studies followed the institution's ethics guidelines and were approved in its ethics process.

\paragraph{Affect Elicitation.}
Participants were exposed to mild discomfort in the form of stress or fear, yet not more so than in daily life.
The stimuli used have been validated in affective psychology and stress research and been found appropriate for the use in experiments with adult participants.

The experiments were conducted in a face-to-face setting to ensure the experimenter could offer aftercare should the participants feel uncomfortable or upset after a session.

\paragraph{Informed Consent and Opt-Out.}
Participants were informed of the requirements (two lab sessions) of the studies in advance.

Participants received a consent form, could ask questions before and during the experiments, and were informed that they could withdraw from the experiment at any time. All participants were able to exercise informed consent.

\paragraph{Deception.}
The participants were deceived in that we did not disclose that our main interest was the password choice. Instead, the personality traits, affect and stress measurements were presented as part of a personality profile Web site.

Participants received a debriefing in which the true purpose of the experiment was explained.

\paragraph{Compensation.}
Participants were reimbursed for their time spent in the experiment at the institution's customary rate for lab experiments. We set the policy that participants would be reimbursed even if they chose to withdraw from the study.

\paragraph{Data Protection.}
We ensured data protection and privacy of the participants' sensitive information. Records were anonymized and stored on an encrypted hard disk. The \Zxcvbn metrics on the participants' passwords were computed offline.

\subsection{Study 1: Effect of Fear}
\subsubsection{Participants}
The participants were recruited through e-mailing lists with in the university on students in computer science, mathematics, and statistics as well as flyers.
The sample consisted of 25 students and 9 participants from a range of professions, incl.  nursing, teaching, and management.

A total sample of $N_\const{F} = 34$ was recruited, 11 women and 23 men. We note that for mood induction procedures, gender has not been found a statistically significant confounding variable~\cite{westermann1996relative}.

88\% of the participants were Caucasian, 9\% from Pacific or Asian islands, one of mixed Asian/Caucasian ethnicity. The majority of participants had a BSc degree (79\%), 12\% a high school degree, 3\% an MSc and 9\% a PhD. Table~\ref{tab:demo_fear} shows the distribution of gender and age ($M = 29.29$, $\vari{SD} = 14.82$).

The sample size of the experiment was determined \emph{a priori} with a sensitivity to detect medium differences between dependent means of Cohen's $d = 0.5$ at $80\%$ power.

\begin{table}[tb]
\centering
\footnotesize
\caption{Demographics of Study 1: Fear}
\label{tab:demo_fear}
\begin{subtable}{.49\columnwidth}
\centering
\begin{tabular}{lr}
\toprule
\textbf{Gender}\\
Female & 32\%\\
Male & 68\%\\
\bottomrule
\end{tabular}
\end{subtable}%
\begin{subtable}{.49\columnwidth}
\centering
\begin{tabular}{lr}
\toprule
\textbf{Age}\\
18--29 & 79\% \\
30--44 & 3\%\\
45--59 & 9\%\\
60+ & 9\%\\
\bottomrule
\end{tabular}
\end{subtable}
\end{table}

\subsubsection{Procedure}
\begin{figure}
\includegraphics[keepaspectratio, width=\columnwidth]{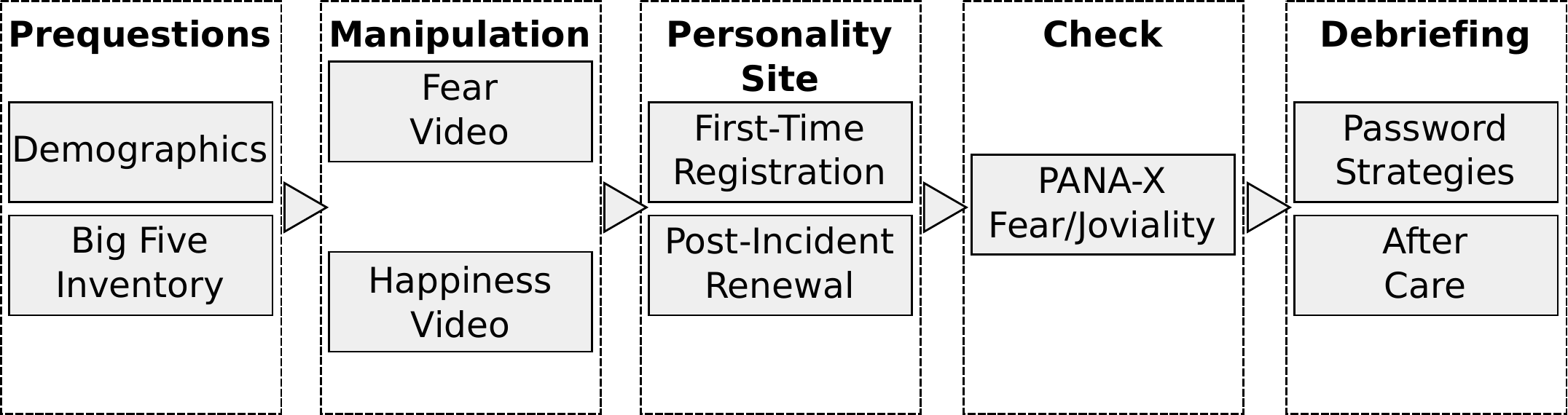}
\caption{Experiment design of Study 1: Fear.}
\label{fig:design_fear}
\end{figure}

We offer an overview of the exact procedure in Figure~\ref{fig:design_fear}.
Participants were constrained randomly assigned to be exposed to either the fear or the happiness stimulus in the first session.

\subsubsection{Manipulation}
Participants were asked to watch standardized stimulus videos that induce either happiness or fear~\cite{ray2007emotion} from the \emph{Handbook of emotion elicitation and assessment}~\cite{coan2007handbook}.
To elicit fear, we selected a specified scene from the \emph{Silence of the Lambs}, in which an FBI agent is trying to find a psychopath, the movie's villain, in a dark basement.

To elicit happiness, we selected a specified scene from \emph{When Harry Met Sally}, in which the two main characters are sitting in a caf{\'e} talking about a fake orgasm and Sally starts to fake an orgasm to prove her point.

The effectiveness of both stimuli has been documented in the corresponding research in affective psychology~\cite{ray2007emotion}.

\subsubsection{Manipulation Check}
The manipulation check allowed us to test whether the manipulation was successful as well as to compute a correlation between the measured strength of the affect and the password strength.

As post-task manipulation check, we administered the \emph{Positive and Negative Affect Schedule} (PANAS-X)~\cite{watson1999panas} as a self-report questionnaire to evaluate the participants' current affects.

The PANAS-X is scale is based on 5-point Likert-items. anchored on 1 -- ``very slightly or not at all,'' 2 -- ``a little,'' 3 -- ``moderately,''   4 -- ``quite a bit,'' and 5 -- ``extremely.''

We restricted PANAS-X to the items pertaining to the variables \textsf{na} (negative affect), \textsf{pa} (positive affect), \textsf{fear} and \textsf{joviality}, which yields a 40-item questionnaire. We anchored PANAS-X on ``at the present moment.''

\subsection{Study 2: Effect of Stress}

\subsubsection{Participants}
The $N_\const{S} = 50$ participants were recruited from university students, mostly with computer science background.
Table~\ref{tab:demo_stress} includes their gender and age distribution ($M = 20.92$, $\vari{SD} = 1.05$).

In terms of gender, the sample was was roughly balanced, with 26 women and 24 men.
We observe that gender is relevant for stress induction procedures. Research of, e.g., Kelly et al.~\cite{kelly2008sex} showed that ``women are more likely than men to report higher levels of negative affect and fear to social stress challenges,'' while they ``do not significantly differ in the extent of autonomic arousal and cortisol reactivity.'' 

The sample size for the stress experiment was determined \emph{a priori} with a sensitivity to detect differences between means of dependent groups of Cohen's $d = 0.4$ with $80\%$ power. This target effect size was chosen to be smaller than the effects observed by Gro{\ss} et al.~\cite{GrCoAl2016} to analyze for incidental stress as alternative explanation.

\begin{table}[tb]
\centering
\footnotesize
\caption{Demographics of Study 2: Stress}
\label{tab:demo_stress}
\begin{subtable}{.49\columnwidth}
\centering
\begin{tabular}{lr}
\toprule
\textbf{Gender}\\
Female & 52\%\\
Male & 48\%\\
\bottomrule
\end{tabular}
\end{subtable}%
\begin{subtable}{.49\columnwidth}
\centering
\begin{tabular}{lr}
\toprule
\textbf{Age}\\
18--29 & 100\% \\
30--44 & 0\%\\
45--59 & 0\%\\
60+ & 0\%\\
\bottomrule
\end{tabular}
\end{subtable}
\end{table}

\subsubsection{Procedure}
\begin{figure}
\includegraphics[keepaspectratio, width=\columnwidth]{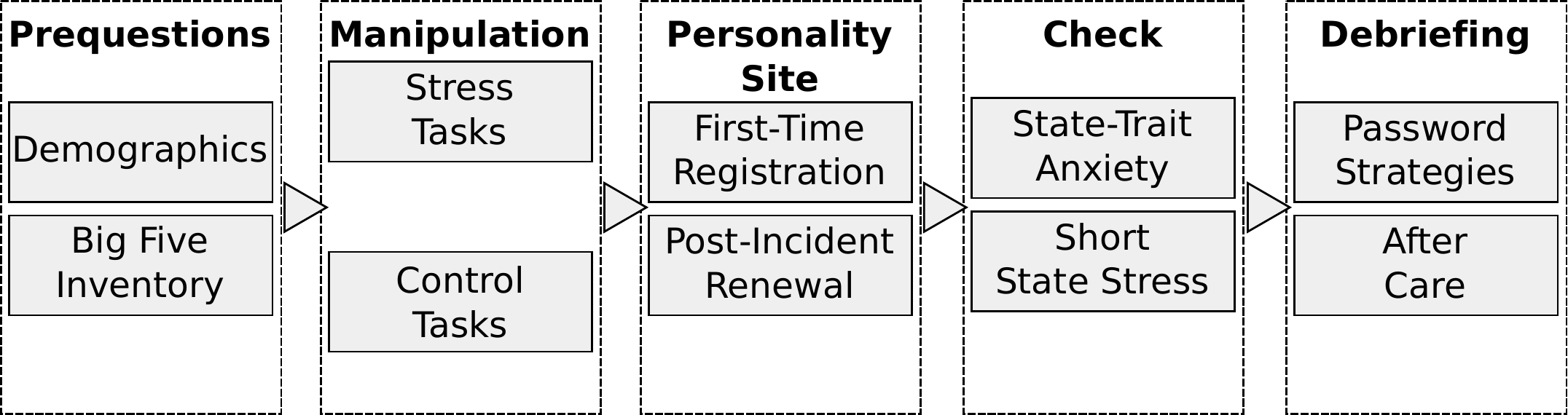}
\caption{Experiment design of Study 1: Stress.}
\label{fig:design_stress}
\end{figure}

Figure~\ref{fig:design_stress} depicts the experiment design and procedure for the second study on stress.
Participants were constrained randomly assigned to be exposed to the stress condition either in the first or the second session.

\subsubsection{Manipulation}
The manipulation in the experiment condition consisted of two stressful tasks combined with induction of social stress. In the control condition, the participants completed balanced tasks which did not induce stress.

\paragraph{Serial Subtraction Task.}
We induced cognitive stress by the \emph{serial subtraction task}, one of the most used tasks to induce psychological stress and receive a cardiovascular response~\cite{li2015laboratory}. In the experiment condition, the participant is asked to continually subtract an one- or two-digit prime number from a four digit number. The participants completed two serial subtraction tasks with 7 and 13 counting down from 9095 and 5245, respectively. Each task lasted 1.5 min. When a participant miscomputed a value, the participant was asked to start over. This was framed as a test of cognitive ability.

In the control condition, participants were asked to continuously add 7 or 13 to 9095 and 5245, respectively, for 1.5 min. If they made a mistake, they were told the correct result, but not to start over.

\paragraph{Isometric Handgrip Task.}
The \emph{isometric handgrip task} has been used to induce physical stress in terms of cardiovascular response~\cite{matthews1988influences}. An electronic hand dynamometer was used to measure the participant's maximal grip strength.

In the experiment condition, the participants were asked to hold the grip at least at 30\% of their maximal grip strength for 2.5 min. Should they go under $30\%$, the experimenter would sternly tell them to keep it above ``Keep it above [30\% of their max].'' The experimenter took notes of the participants's performance. 

In the control condition, the participants were asked to hold the grip at 30\% max strength till they start to feel uncomfortable and, then, to stop. No notes were taken in the control condition.

\paragraph{Social Stress.}
Part of the protocol was to induce social stress in the experiment condition akin to the methods of the \emph{Trier Social Stress Test} (TSST)~\cite{kirschbaum1993trier}. While we did not replicate the TSST exactly, we took inspiration from it. In the serial subtraction task, participants were told that their results would be reviewed with the principal investigator of the study. TSST also used serial subtraction, and similar to the TSST, participants were asked to start over, when they made a mistake.

In the isometric handgrip task's experiment condition, the experimenter was standing behind the sitting participants. The experimenter issued stern warning and made notes whenever the participant slipped under 30\% of the max grip strength.

\subsubsection{Manipulation Check}
We used two instruments to check the stress of participants: the Short State Stress Questionnaire~\cite{helton2004validation,helton2015short} and the State-Trait Anxiety Inventory~\cite{spielberger1970manual}.

\paragraph{Short State Stress Questionnaire.} 
The Short State Stress Questionnaire (SSSQ)~\cite{helton2015short} is self-report questionnaire on state stress, an abridged version of the Dundee State Stress Questionnaire (DSSQ)~\cite{matthews1999validation}. The short version contains 24 questions, formalized as 5-point Likert items anchored on 1 -- ``Strongly Agree,'' 2 -- ``Agree,'' 3 -- ``Neither Agree nor Disagree,'' 4 -- ``Disagree,'' and 5 -- ``Strongly Disagree.'' Questions referred to the time ``in this moment.''

The SSSQ contains three factors: \textsf{engagement}, \textsf{distress}, and \textsf{worry}. We considered the overall \textsf{stress}, the sum of these three, as well as \textsf{distress}.

\paragraph{State-Trait Anxiety Inventory.}
The State-Trait Anxiety Inventory for Adults (STAI-AD)~\cite{spielberger1970manual} is a 40-question self-report questionnaire.
We are interested in the temporary construct of state anxiety, that is, ``how you feel right now.''
It uses 4-point Likert items anchored on 1 -- ``Not At All,'' 2 -- ``Somewhat,'' 3 -- ``Moderately So,'' and 4 -- ``Very Much So.''

\paragraph{NASA Task Load Index.}
The NASA Task Load Index (TLX)~\cite{hart1988development,hart2006nasa} is a standardized and validated instrument to measure the overall task load (strongly related to cognitive effort). It includes sub-scales for mental, physical and temporal demand as well as performance, effort and frustration level.
It measures these sub-scales on visual analogue scales (VAS), which we projected on the interval $[-10, +10]$. For the overall TLX measurement, the different sub-scales are weighted by the subjective workload ranks.

\subsubsection{Debriefing Questionnaire}
The debriefing of Study 2 inquired on multiple aspects of the participants' password choice for both conditions. The questions included:
\begin{compactitem}
  \item \textsf{reuse}: Has the password or a variant been used previously?
  \item \textsf{frequency}: How often has the password been used in the past?
  \item \textsf{last\_use}: When has the password been last used?
  \item \textsf{strategy}: Why has the password been chosen that way?
\end{compactitem}

\section{Results}
As a general rule, statistics were computed with a significance level of $\alpha = .05$. We used two-tailed dependent-samples tests throughout. We consider the manipulation checks of each study as a test family and report $p$-values with Bonferroni-Holm multiple-comparisons corrections as $p_{\const{MC}(n)}$, where $n$ is the number of comparisons made. We do not correct the password strength comparison or the order-effect analysis to ward against Type-II errors.

\subsection{Study 1: Effect of Fear}
\subsubsection{Descriptives}

The study measured PANAS-X \textsf{fear} and \textsf{joviality} as manipulation checks and \Zxcvbn \const{log10} guesses as dependent variable. Table~\ref{tab:desc_fear} offers an overview of the descriptive statistics of the fear experiment ($N_\const{F} = 34$).\tgr{Table cross-checked with Knitr. Transform to automatically generated table?!}

\begin{table}[htb]
\footnotesize
\caption{Descriptive statistics of Study 1: Fear}
\label{tab:desc_fear}
\begin{subtable}{\columnwidth}
\centering
\footnotesize
\caption{Elicited Affect: Fear}
\begin{tabular}{lccc}
  \toprule
 & \multicolumn{2}{c}{PANAS-X} & \Zxcvbn\\
    \cmidrule(lr){2-3} \cmidrule(lr){4-4}
 & \textsf{fear} & \textsf{joviality} & \const{log10} Guesses \\ 
  \midrule
$M$ & 2.91 & 2.35 & 6.39 \\ 
  $\vari{SD}$ & 0.68 & 0.88 & 2.89 \\ 
   \bottomrule
\end{tabular}
\end{subtable}

\vspace{.5cm}

\begin{subtable}{\columnwidth}
\centering
\footnotesize
\caption{Elicited Affect: Happiness}
\begin{tabular}{lccc}
  \toprule
   & \multicolumn{2}{c}{PANAS-X} & \Zxcvbn\\
    \cmidrule(lr){2-3} \cmidrule(lr){4-4}
 &  \textsf{fear} & \textsf{joviality} & \const{log10} Guesses  \\ 
  \midrule
$M$ & 1.12 & 3.47 &  6.74 \\ 
  $\vari{SD}$  & 0.17 & 0.83 &  3.28 \\ 
   \bottomrule
\end{tabular}
\end{subtable}
\end{table}

\subsubsection{Manipulation Check: PANAS-X}
\begin{figure*}[htbp]
\centering
\begin{subfigure}{.33\textwidth}
\centering
\includegraphics[keepaspectratio,width=\textwidth]{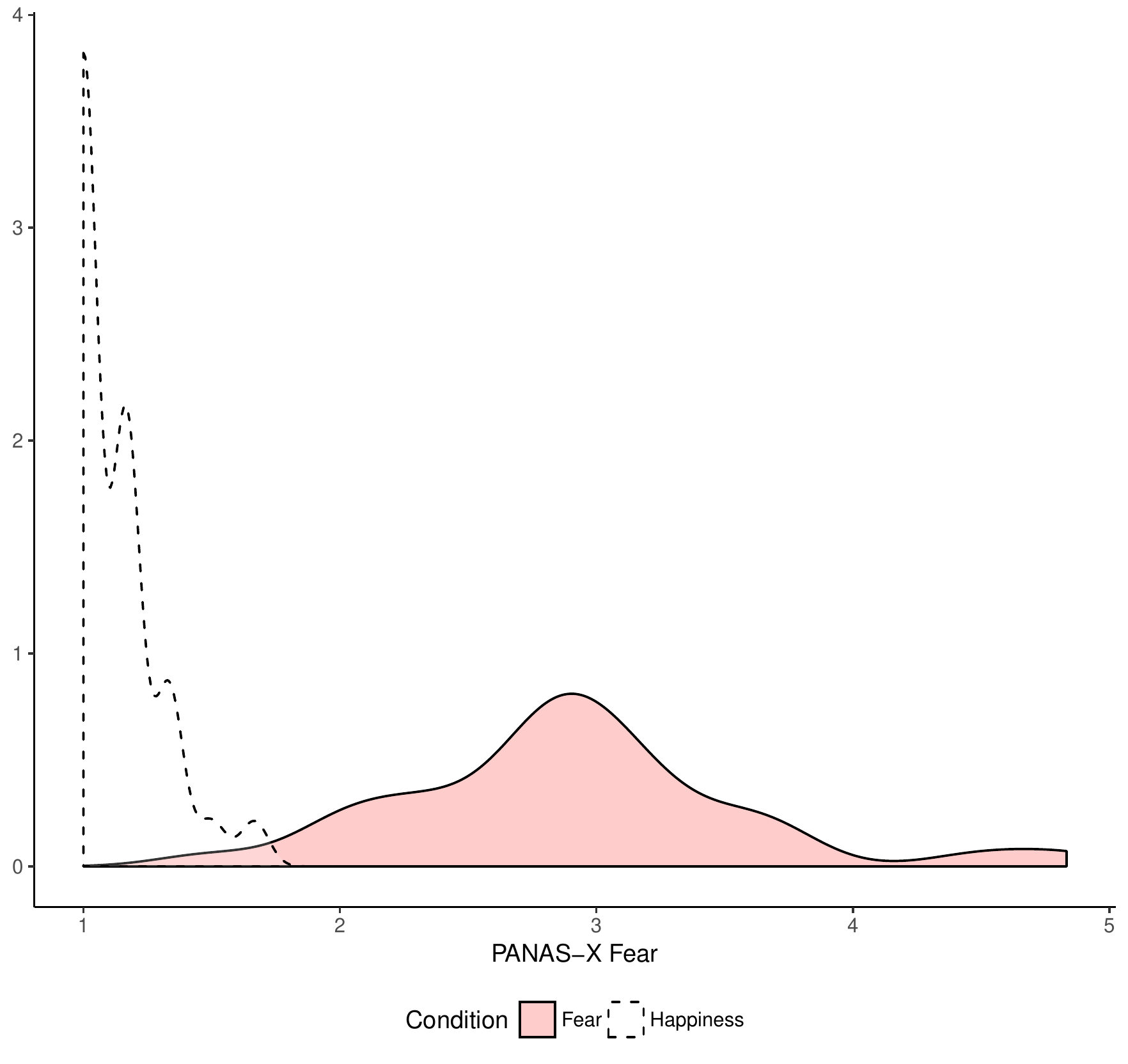}
\caption{PANAS-X \textsf{fear}}
\end{subfigure}~
\begin{subfigure}{.33\textwidth}
\centering
\includegraphics[keepaspectratio,width=\textwidth]{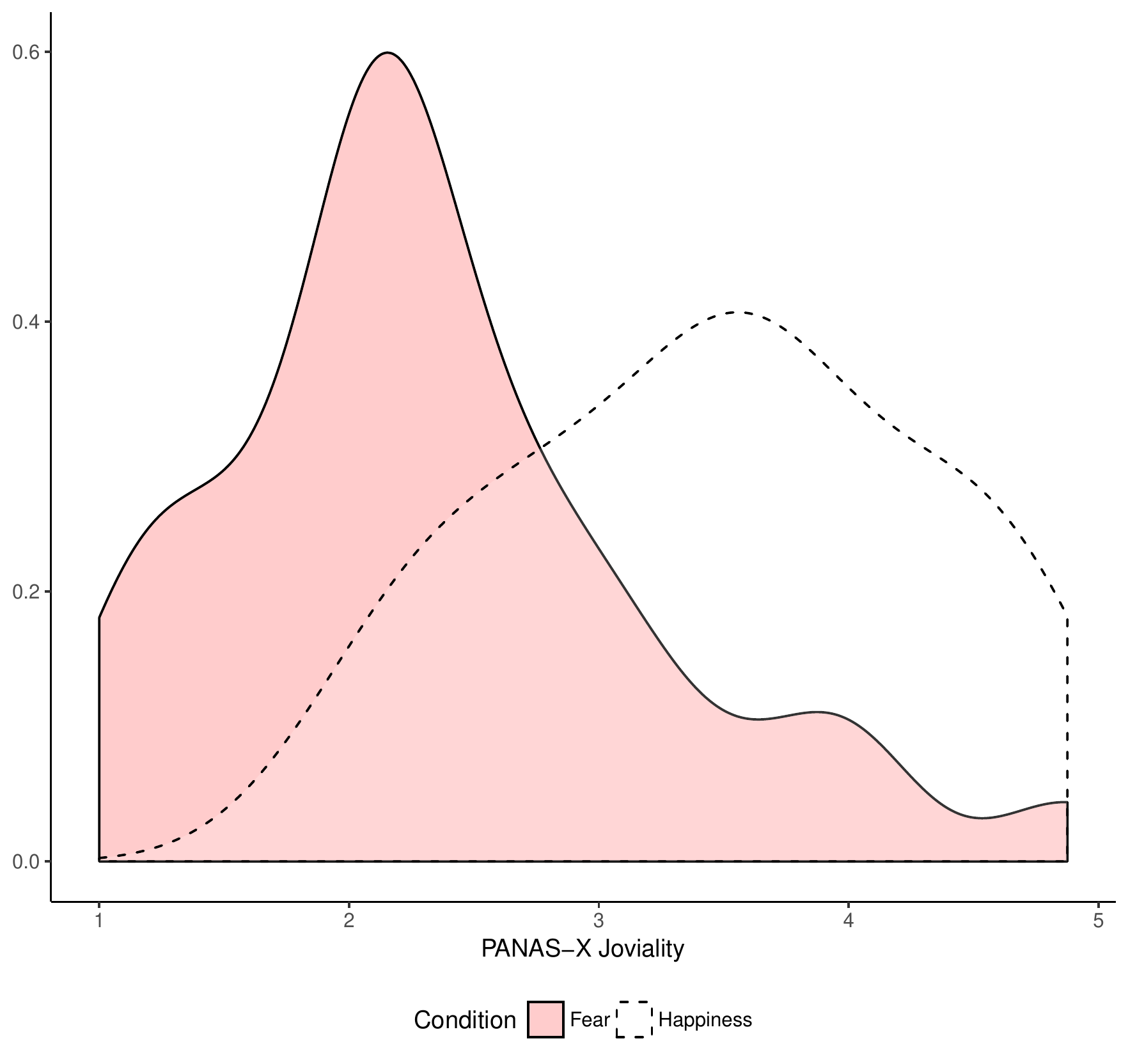}
\caption{PANAS-X \textsf{joviality}}
\end{subfigure}
\caption{Density plots by manipulation check for Study 1: Fear.}
\label{fig:density_mc_fear}
\end{figure*}

As manipulation check, we compared PANAS-X measurements on \textsf{fear} and \textsf{joviality} across the two conditions Fear and Happiness. 
Figure~\ref{fig:density_mc_fear} compares the distributions of the treatments for each measurement.

\paragraph{Assumptions.}
We analyzed the difference of both conditions for outliers based on the Outlier Labeling Rule. 
We further checked for outliers with the Mahalanobis Distance $D^2$ and, finally, concluded that no outlier needed to be capped or removed.

We tested the distribution of differences between conditions for normality with Shapiro-Wilk. While the differences of the \textsf{joviality} measurements were sufficiently normally distributed ($W = 0.96, p = .199$), we found that the differences of the \textsf{fear} measurements were not normally distributed, $W = 0.94, p = .049$.

While the dependent-samples $t$-test is deemed to some extent robust against violations of normality, we complement it with a Wilcoxon Signed-Rank test.

\paragraph{Success of the Fear/Happiness Manipulations.}
Comparing across conditions, the mean \textsf{fear} was statistically significantly greater under elicited fear than under elicited happiness, $t(33) = 15.79, p_{\const{MC}(2)} < .001$, Hedges' $g_{\const{av}} = 4.15$, 95\% CI $[2.64, 4.61]$. We observed a very large effect.

The Wilcoxon signed-rank test confirmed this result, $V = 595, p < .001$.

Comparing across conditions, the mean \textsf{joviality} was statistically significantly less under elicited fear than under elicited happiness, $t(33) = 15.79, p_{\const{MC}(2)} < .001$, Hedges' $g_{\const{av}} = 1.38$, 95\% CI $[0.91, 1.9]$ This was a large effect.

We rejected the null hypothesis $\const{H}_{\const{mc,F,0}}$.
Consequently, the manipulation check showed that the stimulus videos \emph{The Silence of the Lambs} and \emph{When Harry met Sally} indeed caused fear and happiness in the participants.

\esFear

\subsubsection{Password Strength}
We compared the password strength measured in \Zxcvbn \const{log10} guesses across conditions.

\paragraph{Assumptions.}
By the Outlier Labeling Rule and the evaluation of Mahalanobis distance $D^2$, there were no significant outliers.

\paragraph{Differences Between Conditions.}
The \const{log10} number of guesses is not statistically significant across conditions, $t(33) = -0.65, p = .520$, Hedges' $g_{\const{av}} = -0.11$, 95\% CI $[-0.45, 0.23]$. The effect size was trivial. We failed to reject the null hypothesis $\const{H}_{\const{F,0}}$.

The \const{log10} guesses are statistically significantly correlated across conditions, $r = .50$, 95\% CI $[.19, .71]$.

Finally, we conducted a comparison of effect sizes across conditions an the estimation of their confidence intervals.
Figure~\ref{fig:esFear} offers a forest plot of these parameter and interval estimations.

We observed that even if the manipulations yielded large and very large effects, the effect on password strength was trivial. 
The margin of error on the effect size estimation was less than half standard deviation.

\paragraph{Order Effects.}

Dependent samples $t$-tests of \textsf{fear} and \textsf{joviality} by the order of conditions showed no statistically significant difference, $ps > .45$ and $ps > .25$ respectively. We note that the differences of \textsf{fear} by order were neither normally nor symmetrically distributed, by which we used a dependent-samples Sign Test for the corresponding analysis.

Considering the impact of the order on the password strength is interesting in itself, because in the first password choice participants made an initial account registration, in the second password choice the participants made a password reset after an incident.
The differences of \Zxcvbn \const{log10} guesses fulfilled the assumptions (no outliers, normality) for a dependent-samples $t$-test.

\Zxcvbn \const{log10} guesses were not statistically significantly different by condition order, $t(33) = -1.36, p = .183$, $g_{\const{av}} = 0.28$, 95\% CI $[-0.13, 0.7]$.
The mean password strength for the first password (registration) was $M_{\const{F},\const{1st}} = 6.79$.
The mean password strength for the second password (renewal) was $M_{\const{F},\const{2nd}} = 6.43$.
\tgr{Integrate the means with Knitr.}

\subsubsection{Correlation}
\tgr{Switch table to Knitr as well.}
We found that the dependent variable  \Zxcvbn \const{log10} guesses was not statistically significantly correlated with either fear or joviality. Table~\ref{tab:corr_fear} offers an overview of the pair-wise correlations.

\begin{table}[htb]
\centering
\footnotesize
\caption{Correlations of fear vs. password strength in \Zxcvbn \const{log10} guesses  [95\% Confidence Intervals].}
\label{tab:corr_fear}
\begin{tabular}{lcc}
  \toprule
 & PX \textsf{fear} & PX \textsf{joviality} \\ 
  \midrule
PX \textsf{fear} &  &  \\ 
  PX \textsf{joviality} & -0.52*** [-0.67, -0.32] &  \\ 
  \const{log10} Guesses & -0.03 [-0.27,  0.21] &  0.08 [-0.16,  0.31] \\ 
   \bottomrule
\end{tabular}
\end{table}

\subsection{Study 2: Effect of Stress}
\subsection{Data Preparation}
We have analyzed the data for univariate outliers with the Outlier Labeling Rule as well as multivariate outliers with the Mahalanobis distance $D^2$.
We found two cases with extreme values of \Zxcvbn \const{log10} guesses greater than $16$ and $D^2 > 19$.
We decided to cap the outlying values with the 95th percentile, instead of removing the cases altogether.

\subsubsection{Descriptives}
Table~\ref{tab:desc_stress} shows the means and standard deviations of the stress study.
\tgr{The descriptives were cross-checked with Knitr. Would be better to have a direct Knitr integration.}

\begin{table}[htb]
\centering
\footnotesize
\caption{Descriptive statistics of Study 2: Stress.}
\label{tab:desc_stress}
\begin{subtable}{\columnwidth}
\centering
\footnotesize
\caption{Elicited Affect: Stress}
\begin{tabular}{lcccc}
  \toprule
  & \multicolumn{2}{c}{SQQQ} & STAI & \Zxcvbn\\
  \cmidrule(lr){2-3} \cmidrule(lr){4-4} \cmidrule(lr){5-5}
 & \textsf{stress} & \textsf{distress} & \textsf{state\_anxiety} & \textsf{log10} Guesses\\ 
  \midrule
$M$ & 71.50 & 16.78 &  38.50 & 7.96  \\ 
  $\vari{SD}$ &  8.43 &  5.10 &  10.61 & 2.49  \\ 
   \bottomrule
\end{tabular}
\end{subtable}
\vspace{.5cm}

\begin{subtable}{\columnwidth}
\centering
\footnotesize
\caption{Control}
\begin{tabular}{lcccc}
  \toprule
    & \multicolumn{2}{c}{SQQQ} & STAI & \Zxcvbn\\
  \cmidrule(lr){2-3} \cmidrule(lr){4-4} \cmidrule(lr){5-5}
 & \textsf{stress} & \textsf{distress} & \textsf{state\_anxiety} & \textsf{log10} Guesses\\ 
  \midrule
  $M$ &  65.58 & 12.76 & 31.58 & 7.95 \\ 
  $\vari{SD}$ & 9.15 &  4.04 &  7.07 & 2.35 \\ 
   \bottomrule
\end{tabular}
\end{subtable}
\end{table}

\subsubsection{Manipulation Check: SSSQ and STAI}
\paragraph{Assumptions.}
For the Short State Stress Questionnaire (SSSQ) our analysis did not vouch for the capping of any outliers. The differences of the total stress values were normally distributed, Shapiro-Wilk $W = 0.98, p = .580$.

For SSSQ Distress, we observed two outliers, yet not extreme enough to vouch for capping or removal. The differences between distress scores across conditions were normally distributed, Shapiro-Wilk $W = 0.97, p = .158$.

For State-Trait Anxiety (STAI), there were no outliers. The difference between the state anxiety scores of both conditions were normally distributed, Shapiro-Wilk $W = 0.96, p = .081$.

\begin{figure*}[htbp]
\centering
\begin{subfigure}{.33\textwidth}
\centering
\includegraphics[keepaspectratio,width=\textwidth]{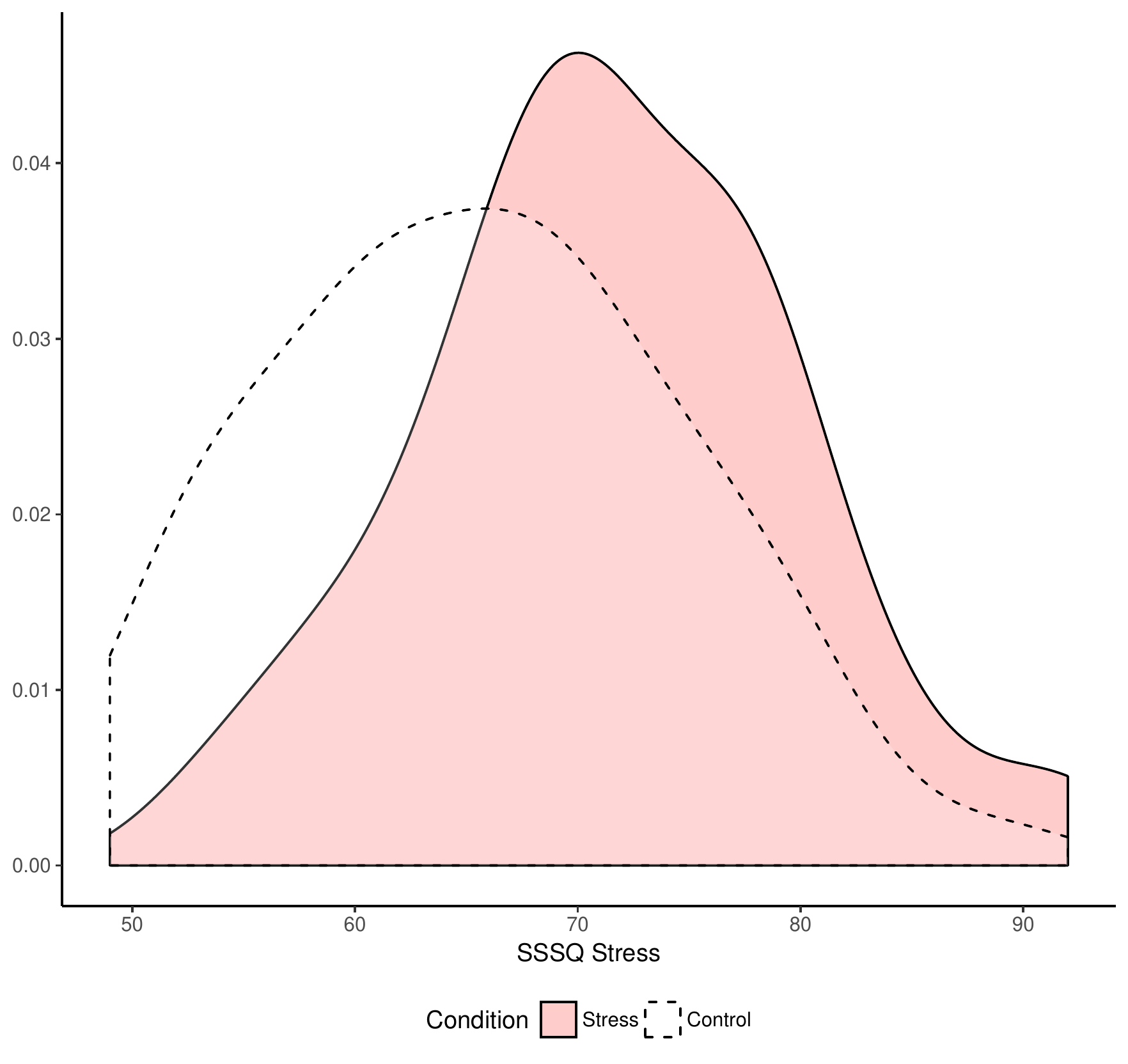}
\caption{SSSQ \textsf{stress}}
\end{subfigure}~
\begin{subfigure}{.33\textwidth}
\centering
\includegraphics[keepaspectratio,width=\textwidth]{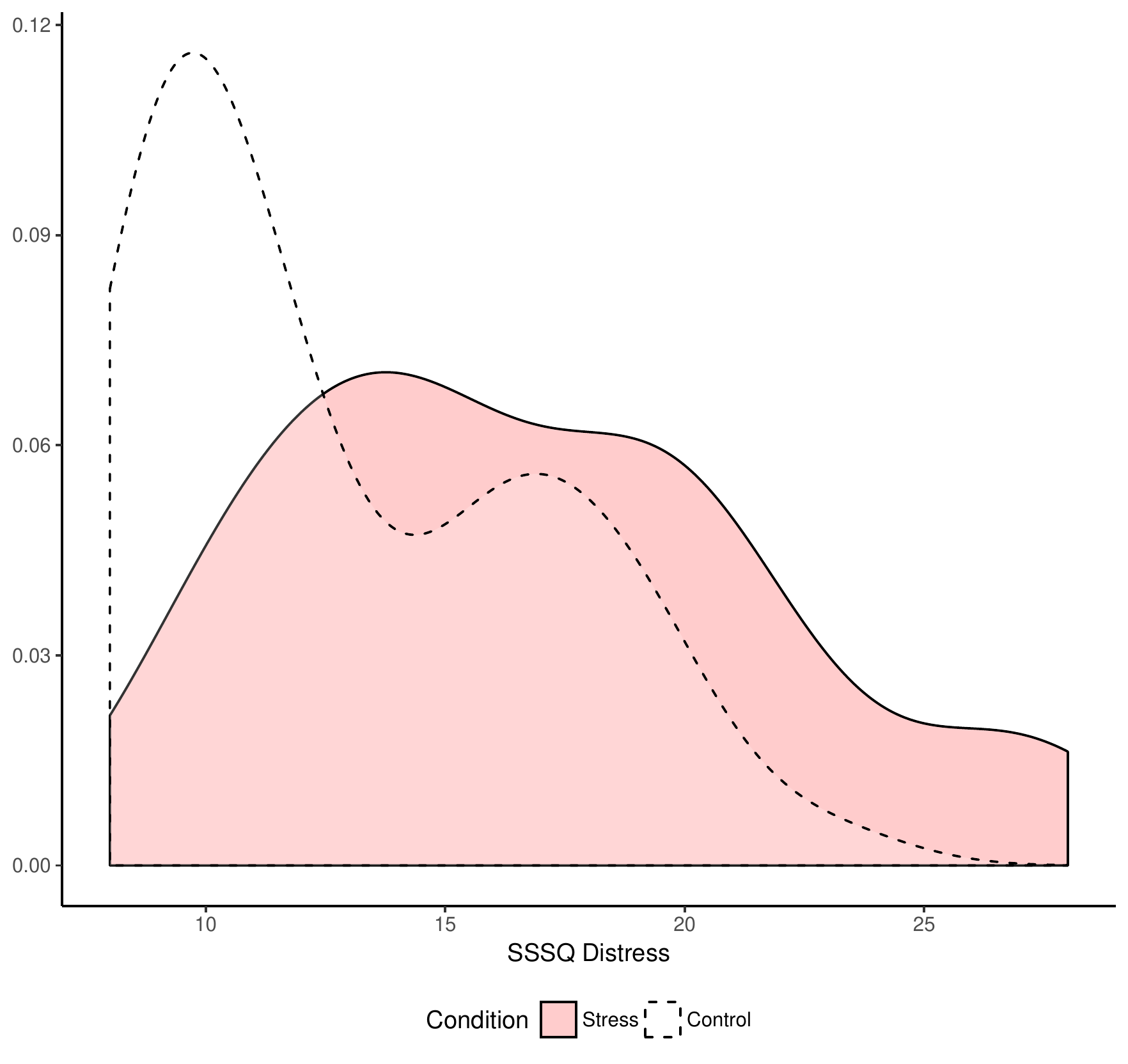}
\caption{SSSQ \textsf{distress}}
\end{subfigure}~
\begin{subfigure}{.33\textwidth}
\centering
\includegraphics[keepaspectratio,width=\textwidth]{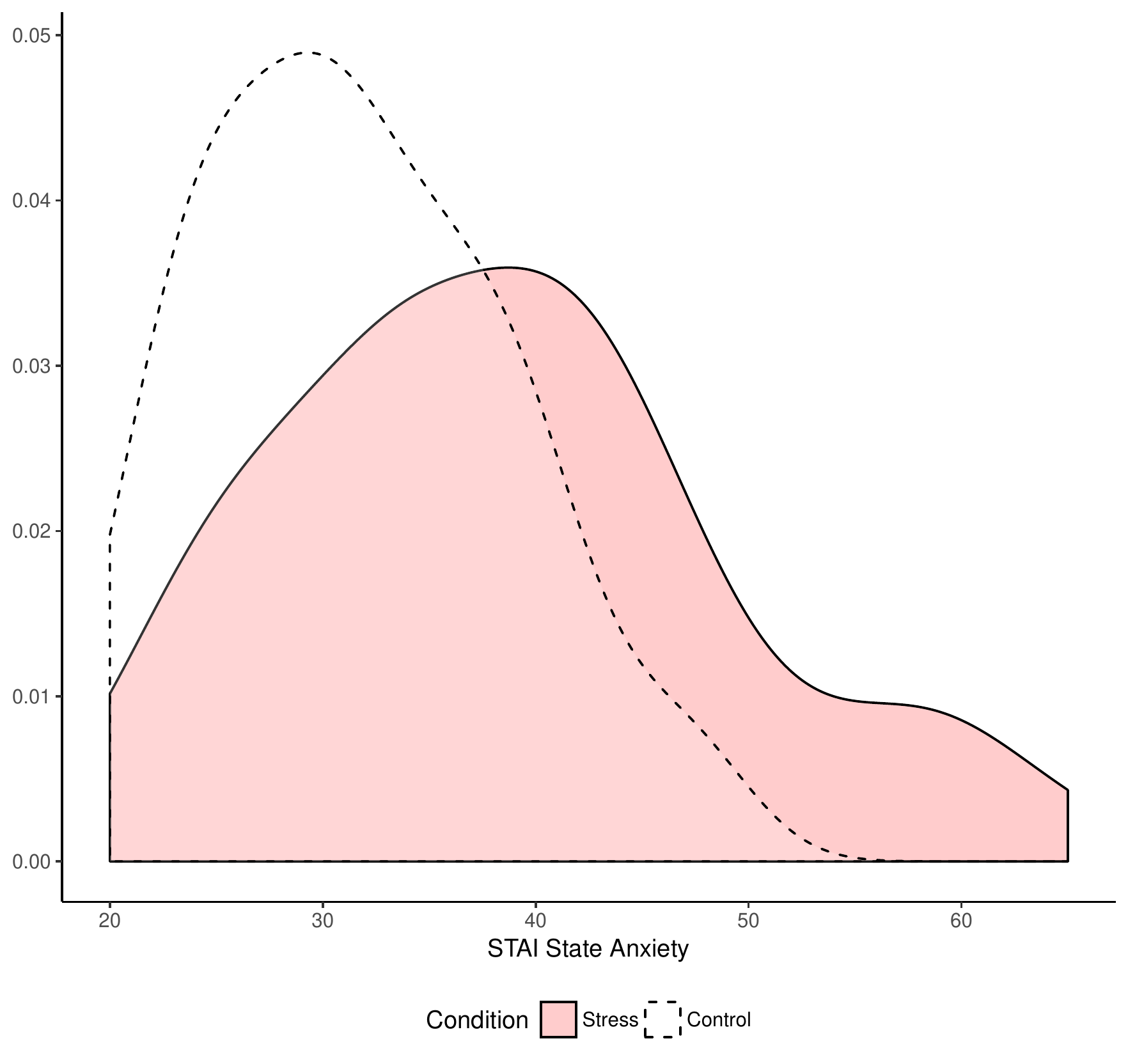}
\caption{STAI \textsf{state\_anxiety}}
\end{subfigure}
\caption{Density plots by manipulation check for Study 2: Stress.}
\label{fig:density_mc_stress}
\end{figure*}

\paragraph{Success of the Stress Manipulation.}
All three measurements, SSSQ \textsf{stress} and \textsf{distress} as well as STAI \textsf{state\_anxiety} showed that the stress manipulations were successful. We offer an overview of the distributions of the treatments by measurements in Figure~\ref{fig:density_mc_stress}.

Participants in the stress condition showed statistically significantly more overall stress than in the control condition, $t(49) = 6.69, p_{\const{MC}(5)} < .001$, Hedges' $g_{\const{av}} = 0.66$, 95\% CI $[0.43, 0.91]$.

They exhibited statistically significantly more distress than in the control condition, $t(49) = 10.78, p_{\const{MC}(5)} < .001$, Hedges' $g_{\const{av}} = 0.87$, 95\% CI $[0.64, 1.11]$. 

Furthermore, they exhibited statistically significantly more state anxiety than in the control condition, $t(49) = 6.46, p_{\const{MC}(5)} < .001$, Hedges' $g_{\const{av}} = 0.88$, 95\% CI $[0.59, 1.17]$. 

As a consequence, we reject the null hypothesis $\const{H}_{\const{mc,S,0}}$.

\subsubsection{Password Strength}
\paragraph{Assumptions.}
While we detected one outlier, we decided not to cap it as it was close to the inner fence. The differences between \Zxcvbn \const{log10} guesses between conditions were normally distributed, $W = 0.96, p = .094$.

\paragraph{Difference Between Conditions}
There was no statistically significant mean difference \const{log10} number of guesses between stress and control condition, $t(49) = 0.04, p = .971$, Hedges' $g_{\const{av}} = 0.01$, 95\% CI $[-0.31, 0.33]$.
Hence, we failed to reject the null hypothesis $\const{H}_{\const{S,0}}$.

There was a statistically significant correlation between \Zxcvbn \const{log10} guesses in the stress and control condition, $r = .33$, 95\% CI $[.06, .56]$.

We include a forest plot of the standardized mean difference of password strength under stress in Figure~\ref{fig:esStress}.
Even though the manipulation caused stress with medium to large effect size, the effect size observed on password strength in \const{log10} guesses is 0, where the confidence interval brackets it to an at most small effect.

\esStress

\paragraph{Order Effects.}
Having checked the assumptions, we computed dependent-samples $t$-tests on \textsf{stress}, \textsf{distress} and \textsf{state\_anxiety} by the order of experiment and control condition.
There were no statistically significant order effects, $ps > . 30$.

In first password choice participants made an initial account registration, in the second password choice the participants made a password reset after a security incident.
\Zxcvbn \const{log10} guesses were not statistically significantly different by condition order, $t(49) = -0.86, p = .397$, $g_{\const{av}} = 0.17$, 95\% CI $[-0.22, 0.56]$.

The mean password strength for the first password (registration) was $M_{\const{S},\const{1st}} = 7.7511819$.
The mean password strength for the second password (renewal) was $M_{\const{S},\const{2nd}} = 8.1579705$.

\paragraph{Difference by Reuse.}
We analyzed the difference in password strength depending on whether participants chose a new password or reused (a variant of) an old one.
Under observation of the corresponding assumptions, computed independent-samples Welch $t$-tests.
The \Zxcvbn \const{log10} guesses were not statistically significantly different by reuse for either the experiment or the control condition respectively, EXP: $t(25.26) = 1.17, p = .253$, Hedges' $g = -0.26$, 95\% CI $[-1.02, 0.5]$ and CTRL: $t(9.34) = -0.66, p = .528$, Hedges' $g = 0.27$, 95\% CI $[-0.49, 1.02]$.

Consequently, we failed to reject the null hypothesis that reuse of passwords has no impact on the password strength.

\subsubsection{Correlation}
We found that there was no statistically significant correlation between the measurements of stress and the password strength in \const{log10} guesses. Table~\ref{tab:corr_stress} contains the overall correlation matrix.

\begin{table*}[htb]
\centering
\footnotesize
\caption{Correlations of stress vs. password strength in \Zxcvbn \const{log10} guesses [95\% Confidence Intervals].}
\label{tab:corr_stress}
\begin{tabular}{lcccc}
  \toprule
 & Total Stress & Distress & Anxiety \\ 
  \midrule
 Total Stress &  &  &  \\ 
  Distress &  0.64*** [ 0.51,  0.75] &  &  \\ 
  State Anxiety &  0.35*** [ 0.16,  0.51] &  0.64*** [ 0.51,  0.74] &  \\ 
  \const{log10} Guesses & -0.02 [-0.21,  0.18] &  0.08 [-0.12,  0.27] &  0.05 [-0.14,  0.25] \\ 
   \bottomrule
\end{tabular}
\end{table*}

\subsection{Stress $\times$ Cognitive Load Interaction}
We observed indications of a disordinal/cross-over interaction between the experiment condition and task load. Figure~\ref{fig:interactConditionTLXMental} offers an interaction diagram illustrating the situation.

\interactConditionTLXMental

We conducted a repeated-measures mixed-effects analysis with TLX Mental Demand and the Experiment Condition as fixed effects. We included a random intercept and distress slope with the subject as the context.

Compared to the intercept model, the model under consideration was on the borderline, but not statistically significant under a Likelihood-Ratio test, $\chi^2(6) = 12.589, p = .050$  $\not< .05$. 

Using the multiple correlations $R^2$ between the original data and the fitted values as an estimate of overall effect size, we obtain $R^2 = 69\%$.
Figure~\ref{fig:assumFitLMEplain} illustrates the fit of the model.

In terms of model diagnostics, we found that the residuals were largely normally distributed, with slight deviations on the tails. We perceived the fitted-residual distribution as largely homoscedastic, even if with a linear bias hinting at a hidden variable.

\assumFitLMEplain

We offer an overview of the model's coefficient estimates in Figure~\ref{fig:interactModelCoef}.

\interactModelCoef

The impact of the Condition $\times$ TLX Mental Demand interaction was statistically significant, $F(1, 47) = 4.868$, $p = .032$,
$0.25$, 95\% CI $[0.03, 0.47]$.

We observe a statistically significant intercept estimate, $F(1, 49) = 899.161$, $p < .001$,
$8.31$, 95\% CI $[7.55, 9.07]$.

The main fixed effects were not statistically significant. Condition: $F(1, 47) = 0.004$, $p = .948$; TLX Mental Demand: $F(1, 47) = 0.261$, $p = .612$.

We observe the following cross-over interaction:
In the control condition, that is when the participants completed non-stressing tasks, participants who reported low mental demand chose passwords with greater mean \const{log10} guesses than participants reporting high mental demand. 

However, in the experiment condition, when the participants completed stressful tasks, participants who reported high mental demand on average chose better passwords than participants who reported low mental demand.

We note that the companion analysis report~\cite{FoGrGr2017_AnaRep} also contains an analysis of the three-way interaction \textsf{Condition} $\times$ \textsf{TLM\_Mental} $\times$ \textsf{Reuse}. The mixed-effects model of the three-way interaction was not statistically significant, $\chi^2(10) = 17.941, p = .056$. Said analysis is congruent with stress and cognitive load having a stronger negative impact on password strength if the user chooses a new password.

\subsection{Meta Analysis of Order Effects}
A meta analysis of the order effects across both studies showed an overall effect in Hedges' $g_\const{av} = 0.219$, 95\% CI $[-0.061, 0.499]$. The summary effect was not statistically significant though, $p = .125$.

Cochran's $Q$-test was not statistically significant at $\alpha = .1$,  $Q(1) = 0.156, p = .693$. We observed a heterogeneity of $I^2 = 0\%$.

\metaAnalysisOrder

\subsection{Network Meta Analysis}
We conducted a network meta analysis to put the effects of different studies in relation.
We considered the 2016 LASER paper by Gro{\ss} et al.~\cite{GrCoAl2016}, which considered the effect of cognitive effort and depletion on password choice.

For the network meta analysis, we coded Gro{\ss} et al.'s categories ``Undepleted,'' ``Effortful'' and ``Depleted'' conditions as three conditions of the same study, noting that, in fact, these categories are derived from the study's manipulation check on depletion level.

We coded the ``Fear'' and ``Happiness'' conditions of Study 1 of this paper, such that ``Happiness'' is mapped onto ``Undepleted.'' Similarly, we coded the ``Stress'' and ``Control'' conditions of Study 2 of this paper, such that ``Control'' is mapped onto ``Undepleted.''

\networkMetaAnalysis

We display an overview of the resulting network meta analysis in Figure~\ref{fig:networkMetaAnalysis}. The meta analysis is based on standard mean differences, Hedges' $g$ in case of Gross et al. and Hedges' $g_\const{av}$ in case of this paper. Figure~\ref{fig:nma_forest} yields a forest plot of the results, while Figure~\ref{fig:nma_network} shows the network of treatment relations.

We observe that the effects associated with the ``Effortful'' and ``Depleted'' categories of Gro{\ss} et al.~\cite{GrCoAl2016} are maintained at large and medium effect sizes. The treatments fear and stress only yield trivial effect sizes. Consequently, stress is not supported as an alternative explanation for reported effects of cognitive effort and depletion.

\section{Discussion}
\subsection{Incidental fear and stress were successfully induced.}
We induced incidental fear and stress, that is, an affective state not related to the password choice scenario. Consistently yielding large to very large effect sizes, the different induction techniques (affect stimulus videos, stress battery) were shown to have worked well.

Clearly, task-related fear and stress should more likely to impact password strength, as pursued in research on fear appeals. For instance, participants could be exposed to a news article describing the negative impact of identity theft or a password breach. Such an experiment setup would vouch for an analysis with the Protection Motivation Theory (PMT)~\cite{rogers1983cognitive,floyd2000meta,witte1992putting}.

\subsection{Incidental fear and stress is likely to have at most a small negative effect on password choice.}
While the 95\% confidence intervals on the effect of fear and stress on password strength bracketed the effects as small, the effect of incidental stress corrected for other predictors was negative, similarly to the effect of incidental fear. Hence, future research may aim at pinpointing a negative influence of incidental stressors.

At the same time, we found a statistically significant interaction between stress and mental demand, which asks for further investigation seeking to isolate both conditions. 

\subsection{Whether reusing  an existing password or creating a new one serves the user better may be situational.}
While a three-way interaction model including password reuse~\cite{FoGrGr2017_AnaRep} was not statistically significant, we observe weak evidence that newly created passwords would be weaker when the user is either stressed or under mental demand, and stronger under baseline conditions.

While recommendations that users should create new passwords when they are rested  as well as be allowed to rely on variants of prior passwords when they are stressed or depleted seem plausible, this area requires further investigation for a conclusive result.

\subsection{Cognitive effort and depletion are maintained as strong effects on password choice.}
Gro{\ss} et al. hypothesized that stress could be an alternative explanation for the observed effects
\begin{inparaenum}[(a)]
\item that users under cognitive effort but not depleted created better passwords than the control group, and
\item that users reporting high depletion create worse passwords than the control group.
\end{inparaenum}

Having induced incidental fear and stress and compared the results in a network meta analysis, we have not found evidence that these factors cause an effect of similar magnitude as the the cognitive effort and depletion, reported in Gross et al.~\cite{GrCoAl2016}. Hence, cognitive effort and depletion are still plausible explanations of the observed differences in password strength.

\subsection{When users are asked to renew their password because of a security incident, their password strength may improve.}
Having considered the meta analysis of differences between password choice in a first-time registration and in a renewal after a security incident, we found evidence of a consistent effect between studies that the password strength in the renewal condition is slightly greater than in the first-time registration.

In terms of magnitude, we find that the greater \const{log10} number of guesses during renewal after a reported security incident makes for a small effect size, $g_\const{av} = 0.219$, 95\% CI $[-0.061, 0.499]$.

This yields an early indication that task-related fear (induced by a security incident message) is causing a change in user behavior towards a protection motivation. Hence, we expect that a future experiment constructed to test the full Protection Motivation Theory (PMT) in a password choice scenario will yield conclusive result.

Given the small effect size estimates, however, neither the individual studies nor the meta analysis had enough power to reject the null hypothesis with statistical significance. Hence, this asks for further investigation. 

\subsection{Limitations}
\subsubsection{Generalizability}
The participants were largely recruited from university students, limiting generalizability.

In terms of ecological validity, the studies created a scenario in which actual private information of participants (personality traits, stress and anxiety data) was stored on a Web site. Having been made aware of the sensitivity of such personal data, the participants' incentive to protect the data was similar to real life.

The participants were exposed to a diversion in that they were only informed after the experiment that the research aim included the password strength and in that they were misled under the pretext of a security incident to change their password. Consequently, the first and the second run of the password choice trial were different by design in terms of ``first registration'' vs. ``password change after incident''.

\subsubsection{Constraints on Bounding Small Effects}
We note that an within-subjects experiment to establish a lower bound on the impact of fear or stress in password strength would need a considerable number of participants. For $95\%$ \emph{a priori} power on a dependent-samples $t$-test for standardized mean difference of Cohen's $d = 0.1$, one would need a sample size of $1300$.

\section{Conclusion}

This is the first work to investigate the effects of incidental fear and stress on password choice. It is the first to estimate the magnitude of such effects across studies on the influence of the user's current cognitive and affective state on password decision making.

As future work, the two studies yield an observation on the effect of fear appeals in password choice. There were first indications that the message of a security incident caused participants to choose a stronger password with a small effect size. This vouches for further investigation, for instance using the full Protection Motivation Theory (PMT)~\cite{rogers1983cognitive,floyd2000meta,witte1992putting} as a foundation.

\section*{Acknowledgements}
We are grateful for the discussions with Kovila Coopamootoo, especially on earlier work on cognitive effort, on incidental and integral affect, as well as on the Protection Motivation Theory. We are grateful for the discussions with Uchechi Phyllis Nwadike, especially on the effects of incidental fear, sadness and happiness on privacy decision making~\cite{NwGrCo2016}. We appreciated discussions with Roy Maxion on experimentation on stress. This work was in parts supported by the \CASCAde. This research was conducted as part of the EPSRC Research Institute in Science of Cyber Security (RISCS II).

\balance

\begin{DocumentVersionSTAST}
\bibliographystyle{ACM-Reference-Format}
\end{DocumentVersionSTAST}
\begin{DocumentVersionTR}
\bibliographystyle{IEEEtran}
\end{DocumentVersionTR}
\bibliography{laser,psychology_affect,psychometry,passwords,stress,methods_resources,fearappeals}

\begin{appendix}




\section{Cross-Correlations}
We determined in Studies 1 and 2 that the \Zxcvbn \const{log10} guesses were not statistically significantly correlated with either fear or stress. Figure~\ref{fig:corrgramFear} and Figure~\ref{fig:corrgramStress} the corresponding corrgram for fear and stress respectively.

\corrgramFear

\corrgramStress

\section{Effect and Interval Estimates}

\esComparisonAllCombined
\end{appendix}

\end{document}